\documentclass[]{iopart}
\usepackage{iopams}
\usepackage{amssymb}
\usepackage{graphicx}
\usepackage{psfig}
\usepackage{portland}

\def\beq{\begin{equation}}
\def\eeq{\end{equation}}
\def\bea{\begin{eqnarray}}
\def\eea{\end{eqnarray}}
\def\mr{\mathrm}
\def\mR{\mathcal{R}}
\def\ug{\,=\,}
\def\lp{\left(}
\def\rp{\right)}
\def\lP{\left[}
\def\rP{\right]}
\def\pr{\prime}
\def\pt{\partial}
\def\disp{\displaystyle}
\def\nnu{\nonumber}
\def\ltord{\hbox{$\;\raise.4ex\hbox{$<$}\kern-.75em\lower.7ex\hbox{$\sim$}
                       \;$}}
\def\gtord{\hbox{$\;\raise.4ex\hbox{$>$}\kern-.75em\lower.7ex\hbox{$\sim$}
                       \;$}}

\begin{document}

\title[Curvature profiles as initial conditions for
primordial black hole formation]{Curvature profiles as initial
conditions for primordial black hole formation}
\author{Alexander G. Polnarev, Ilia Musco}

\address{Astronomy Unit, Queen Mary University
of London, Mile End Road, London E1 4NS, England\\}

\begin{abstract}
This work is part of an ongoing research programme to study
possible Primordial Black Hole (PBH) formation during the
radiation dominated era of the early
universe. Working within spherical symmetry, we specify an initial
configuration in terms of a curvature profile, which represents
initial conditions for the large amplitude metric perturbations,
away from the homogeneous Friedmann Robertson Walker model, which
are required for PBH formation. Using an asymptotic
quasi-homogeneous solution, we relate the curvature profile with
the density and velocity fields, which at an early enough time,
when the length scale of the configuration is much larger than the
cosmological horizon, can be treated as small perturbations of the
background values. We present general analytic solutions for the
density and velocity profiles. These solutions enable us to
consider in a self-consistent way the formation of PBHs in a wide
variety of cosmological situations with the cosmological fluid
being treated as an arbitrary mixture of different components with
different equations of state. We show that the analytical
solutions for the density and velocity profiles as functions of
the initial time are pure growing modes. We then use two different
parametrisation for the curvature profile and follow numerically
the evolution of a range of  initial configuration.
\end{abstract}

\today


\section{Introduction}
\label{intro} The possible existence of primordial black
holes (PBHs) was first proposed in 1966 by Zeldovich \&
Novikov \cite{Zeld_Nov} and, at the beginning of 1970s,
by Hawking \cite{Hawking}. It was then widely discussed
in the following years, see for example the recent review
by Carr for a full list of references \cite{Carr1}. In
1974 Hawking made his famous discovery of black hole
evaporation \cite{Hawking_evap}, that is cosmologically
relevant if the mass of the black holes concerned is less
than $10^{15}$ grams. For this reason the problem of PBH
formation started to be attractive and it was widely
investigated in the following 30 years.  

The PBH formation process was first investigated by Carr
(1975) \cite{Carr2} using a simplified model for an
overdense collapsing region, described as a closed
Friedmann-Robertson-Walker (FRW) universe, surrounded by
a spatially flat FRW expanding background. In the
radiation dominated epoch of the Universe this leads to a
threshold value for the perturbation amplitude
$\delta_c$, where the amplitude $\delta$ is defined as
the mass excess in the overdense region, and the black
holes formed have masses of the order of the horizon mass
at the time of formation. In this way Carr obtained a
first rough estimate of $\delta_c \sim 1/3$ (evaluated at
the time of horizon crossing, when the overdensity region
enters into the cosmological horizon) by comparing the
Jeans length  with the cosmological horizon scale at the
time of black hole formation. Subsequently a self
consistent hydrodynamical analysis of PBH formation was
carried out by Nadezhin, Novikov and Polnarev in 1978
\cite{Nadezhin} and in 1980 \cite{Novikov} using, for the
first time in this context, a hydrodynamical computer
code written in the Misner-Sharp slicing, characterised
by a diagonal metric with a cosmic time coordinate that
reduces to the FRW metric in the absence of
perturbations. Previously the same slicing was used by
Podurets \cite{Podurets} and May \& White \cite{May} to
study stellar core collapse. An alternative analysis was
developed in 1979 by Bicknell \& Henriksen
\cite{Bicknell} using a different method based on
integration along hydrodynamical characteristics. These
papers showed that the threshold amplitude of the
perturbation is dependent on the particular shape of the
initial conditions. It was also found in \cite{Nadezhin}
that pressure gradients considerably reduce the PBH mass
formed at the end of the hydrodynamical process.

In the following twenty years attention was concentrated
on different aspects of PBH formation. For example,
calculating  the amount of Hawking radiation emitted by
sufficiently small PBHs, with a mass less than $10^{15}$
g, one can obtain important constraints on parameters
that characterise the different epochs and processes of
the Universe (see, for example, \cite{Zeldovich,Green1}
and references there). The formation of  PBHs was
considered also within different scenarios, for example
during phase transitions \cite{Jedamzik}, with a soft
equation of state \cite{Khlopov}, by collapse of cosmic
loops \cite{Hawking2}, \cite{Polnarev} or from bubble
collisions \cite{Crawford}. In general, the study of PBHs
provides a unique probe for different areas of physics:
the early Universe, quantum gravity, gravitational
collapse and  high energy physics. This is explained with
full lists of references in the various reviews of PBHs,
see for example \cite{Carr1}. 

More recently, in 1999, Niemeyer and Jedamzik
\cite{Niemeyer2,Niemeyer1} made new numerical
calculations pointing out the relevance of scaling laws
for PBH formation. They showed that the black hole mass
$M_\mr{BH}$ follows a power law
$(\delta-\delta_c)^\gamma$ if $\delta$ is close enough to
$\delta_c$, the same behaviour as seen in critical
collapse by Choptuik \cite{Choptuik} and other authors
(see review \cite{Gundlach}). Niemeyer and  Jedamzik
found that $\delta_c \,\simeq\, 0.7$ for the three types
of perturbation profile that they studied. In the same
year, Shibata and Sasaki \cite{Shibata} presented an
alternative formalism for studying PBH formation focusing
on metric perturbations, rather than density
perturbations, as had also been done previously
\cite{Nadezhin,Ivanov}. They pointed out that the initial
conditions used in \cite{Niemeyer1} were specified
initially within a nonlinear regime of perturbations of
the energy density and velocity field, and were therefore
``inevitably contaminated by an unrealistic decaying
mode'' component that would diverge for
$t\rightarrow0$. In a recent paper \cite{Green3} this
analysis has been converted in terms of the perturbation
amplitude $\delta$, showing that the Shibata and Sasaki
results, corresponding to a wide choice of perturbation
shapes, are consistent with  $\delta_c$ in the range $0.3
\ltord \delta_c \ltord 0.5$.

The disagreement between  this and the value $\delta_c
\,\simeq\, 0.7$ has been explained by Musco \emph{et al}
\cite{Musco}, where simulations similar to those used in
\cite{Niemeyer2} were carried out but specifying the
initial conditions within the linear regime, giving only
growing mode solutions at the horizon crossing time. The
simulations in \cite{Musco} give  values of $\delta_c$ in
the range $0.43\,-\,0.47$ (instead of $0.67\,-\,0.71$)
for the same types of perturbation profiles used in
\cite{Niemeyer1}. 

The present work is a new analysis of PBH formation using
the numerical technique developed in \cite{Musco},
implemented together with a quasi homogeneous solution
(\cite{Lifshits}, \cite{Nadezhin}) that, assuming
spherical symmetry, can be characterised by a single
function of the radial coordinate. In this paper this
function, called $K(r)$, is chosen as a curvature profile
that allows a self consistent determination to be made of
the whole set of initial conditions. This is a further
development of the original approach of Nadezhin, Novikov
and Polnarev \cite{Nadezhin}. 

According to the asymptotic quasi homogeneous solution
\cite{Lifshits}(as $t \rightarrow 0$), an arbitrary
curvature profile $K(r)$ corresponding to a large
amplitude perturbation of the metric does not depend on
time, while perturbations of energy density and velocity
vanish asymptotically as $t \rightarrow 0$. Therefore
these can be treated as small perturbations. Solving the
equations for the small velocity and density
perturbations analytically we impose in a self consistent
way all of the initial conditions corresponding to a
moment in time when the quasi homogeneous solution of a
certain order is valid. The curvature profile $K(r)$
appears as the source in the right hand side of the
relevant equations and we can say that the density and
velocity perturbations which we use as initial conditions
are generated by the curvature $K(r)$. Then we use the
computer code to follow the subsequent non linear
evolution of the initial configuration (by the
configuration we mean the region of strong metric
perturbation). To avoid confusion we should emphasise
that the small perturbations predicted by any
cosmological model are relevant in this context only when
one calculates the probability of finding a configuration
with a high amplitude perturbation of the metric. In
fact, the small perturbations of density and velocity
which we discuss here are determined not by arbitrary
cosmological initial conditions, but entirely from the
curvature profile within the initial configuration.

The paper is organised as follows. 

In section 2 we give a
very brief description of the Misner-Sharp equations. We
then use these equations to specify all of the initial
conditions, which we then use in our numerical
computations, in terms of a curvature profile. When the
length scale of the configuration is much larger than the
cosmological horizon, the Misner-Sharp equations can be
reduced to a system of linear differential equations
which we solve analytically.  

In section 3 we obtain solutions for a rather general
multi-fluid case, where the equation of state corresponds
to an arbitrary mixture of perfect fluids. We demonstrate
that, in the simple case of single perfect fluid,
corresponding perturbations of energy density and
velocity, obtained within the framework of the quasi
homogeneous solution, evolve with time like a pure
growing mode in standard cosmological perturbation
theory, while their space dependence is entirely
determined by the curvature profile $K(r)$. 

In section 4 we discuss the physical properties of the
curvature profile $K(r)$, which is directly connected to
the time-independent comoving curvature perturbation
$\mR$ frequently used in literature.

In section 5 we introduce two different parametrisation of
$K(r)$.

In section 6, we present numerical tests to demonstrate
self-consistency of the initial conditions. We then show
that the initial conditions described by the quasi
homogeneous solution have been imposed consistently in
the code and give numerical examples of primordial black
hole formation in the case of a radiation dominated
universe. We show how the threshold for black hole
formation is linked with the curvature profile and
discuss the results obtained with the two different
parametrisation used in the computations.

Summary and conclusions are given in section 7.

\section{Mathematical formulation of the problem}

\subsection{The Misner-Sharp equations}
Assuming spherical symmetry, it is convenient to divide
the collapsing matter into a system of concentric
spherical shells and to label each shell with a
Lagrangian co-moving radial coordinate which we denote as
$r$. Then the  metric can  be written in the form used by
Misner \& Sharp \cite{Misner}  
\beq
ds^2=-a^2\,dt^2+b^2\,dr^2+R^2\lp d\theta^2+\sin^2\theta
d\varphi^2\rp, \label{sph_metric} \eeq 
where $R$ is a circumference coordinate, $a$ and $b$ are
functions of $r$ and of the time coordinate $t$ , which
reduces to the familiar FRW time coordinate (referred to
in literature as ``cosmic time'') in the absence of
perturbations. The FRW metric used to describe
homogeneous and isotropic cosmological models  is a
particular case of (\ref{sph_metric}): 
\beq 
ds^2 \ug
- dt^2 + s^2(t) \lP \frac{dr^2}{1-Kr^2} + r^2\lp
d\theta^2+\sin^2\theta d\phi^2\rp\rP \label{FRW} 
\eeq 
where $s(t)$ is the scale factor and $K$ is the curvature
parameter that is equal to $0$, $+1$ and $-1$ for flat,
closed and open universes. 

For a classical fluid, composed of particles with nonzero
rest-mass, it is convenient to use the rest-mass $\mu$
contained interior to the surface of a shell (or,
equivalently, the baryon number) as its co-moving
coordinate $r$. For fluids not possessing these conserved
quantities, one can still define a ``relative compression
factor'' $\rho$ \cite{Miller2,Miller3} (equivalent to the
rest-mass density on a non relativistic fluid) and one
then has 
\beq 
d\mu=4\pi\rho R^2b\,dr\,. \label{d_mu} \eeq Identifying
$\mu$ and $r$ one has \beq b=\frac{1}{4\pi R^2\rho}. 
\eeq
Following the notation of \cite{Misner}, we write the
equations in terms of the operators 
\beq 
D_t\equiv\frac{1}{a}\lp
\frac{\partial}{\partial t}\rp, \label{D_t} 
\eeq 
\beq
D_r\equiv\frac{1}{b}\lp \frac{\partial}{\partial\mu}\rp,
\label{D_r} 
\eeq 
and define 
\beq U \equiv D_t R, \label{U} 
\eeq
\beq \Gamma \equiv D_r R, \label{Gamma1} 
\eeq 
where $U$ is the radial component of the four-velocity in
the associated Eulerian frame, $R$ is the circumference 
coordinate, and $\Gamma$ is a generalization of the
Lorentz factor. 

We assume that the matter consists of different
components and that each component can be described as a
perfect fluid with the equation of state 
\beq 
p=\gamma\,e, \label{eq.state} 
\eeq 
where $p$ is the pressure and $e$ is the energy
density. For any equation of state of the form $p=p(e)$,
the system of Einstein equations can be written as:
\beq 
D_tU=-\lP \frac{\Gamma}{(e+p)}D_rp+\frac{M}{R^2}+4\pi
Rp\rP, \label{Euler1} 
\eeq 
\beq D_t\rho=-\frac{\rho}{\Gamma
R^2}D_r(R^2U), \label{D_trho} 
\eeq 
\beq D_t e=\frac{e+p}{\rho}D_t\rho,\label{D_te} 
\eeq 
\beq D_t M=4\pi pUR^2,
\label{D_tM} 
\eeq 
\beq 
D_r a=-\frac{a}{e+p}D_r p,\label{D_ra} 
\eeq
\beq D_r M=4\pi\Gamma eR^2, \label{D_rM} 
\eeq 
where $M$ is a measure of the mass-energy contained
inside radius $\mu$ and $\Gamma$ can be calculated either
from (\ref{Gamma1}) or from the constraint equation 
\beq 
\Gamma^2=1+U^2-\frac{2M}{R}
\label{Gamma}. 
\eeq

When the cosmic time approach is used for calculations of
gravitational collapse leading to black hole formation it
has the well-known drawback that singularities appear
after a finite time in the calculations, before an event
horizon has formed. Once a singularity has formed, the
calculations cannot be continued further and so parts of
the evolution which could potentially be seen by an
outside observer cannot be followed in this way. In
particular, it is not possible to follow all of the
process of the formation of the event horizon. This
drawback is the reason for an observer-time formulation
being used in the present work rather than continuing to
use the cosmic time formulation. Hernandez \& Misner
\cite{Hernandez} introduced the concept of ``observer
time'', using as the time coordinate the time at which an
outgoing radial light ray emanating from an event reaches
a distant observer. In the original formulation, this
observer was placed at future null infinity but for
calculations in an expanding cosmological background we
use an FRW fundamental observer sufficiently far from the
perturbed region so as to be unaffected by the
perturbation. A complete description of the
Hernandez-Misner equations was given in \cite{Musco}. 

\subsection{Perturbations in the quasi homogeneous solution}
To describe PBH formation, we follow the quasi
homogeneous approach (\cite{Nadezhin}, \cite{Novikov})
and solve the Misner-Sharp system of equations, giving a
non perturbative description of large amplitude metric
perturbations away from the homogeneous FRW model. Using
this approach, as mentioned in the introduction, we can
avoid the unphysical arbitrariness in the choice of
initial conditions considered in previous works
\cite{Niemeyer2,Shibata,Musco}. 

The characteristic feature of the asymptotic quasi
homogeneous solution is that all mass elements expand
asymptotically, when $t\rightarrow 0$, according to the
FRW model, with energy-density $e\,=\,1/6(1+\gamma)^2 \pi
t$ for an equation of state given by (\ref{eq.state}),
while a spatial hypersurface with $t\,=\,const$ can have
arbitrary curvature if the perturbation has a
length-scale sufficiently larger than the cosmological
horizon. As a consequence of this the hydrodynamical
quantities can be considered as small perturbations with
respect to the background solution, while the curvature
perturbation is arbitrary large. This curvature
perturbation is time independent, because pressure
gradients in this regime are negligible (\cite{Lifshits},
\cite{Nadezhin}, see also, for example,  \cite{Liddle}).

In this work, we consider spherical symmetry and specify
conditions on an initial space-like hypersurface using a
time independent function $K\,=\,K(r)$ that represents
the curvature profile. As we will see in the next
section, this $K(r)$ is directly related to the comoving
curvature perturbation $\mR$. We can also see that
introducing a profile $K(r)$ into the Friedmann Robertson
Walker metric (\ref{FRW}) one obtains an asymptotic
solution of the Einstein equation in the limit $t
\rightarrow 0$. We therefore use $K(r)$ to represent
initial curvature perturbations with scales much larger
than the horizon, and solve the system of perturbed
differential equations to express the set of
hydrodynamical variables in terms of $K(r)$. 

The system of equations (\ref{U}) - (\ref{Gamma}) can be
re-written as:  
\bea
& \dot R = aU \label{R_dot}\\
& \frac{\dot b}{b} = a\frac{U^\pr}{R^\pr}
\label{b_dot} \\
& \frac{a^\pr}{a} = -\frac{\gamma}{1+\gamma}\frac{e^\pr}{e}
\label{a_prime} \\
& \dot M = -4\pi \gamma eR^2\dot R \label{M_dot} \\
& M^\pr = 4\pi eR^2R^\pr \label{M_prime} \\
& \frac{R^{\pr 2}}{b^2} = 1 + U^2 - \frac{2M}{R} \label{R_prime}
\eea 
where the dot and dash denote differentiation with
respect to $t$ and $r$ respectively, we use the equation
of state (\ref{eq.state}) to express the pressure as a
function of the energy density, equation (\ref{Gamma1})
gives an expression for $\Gamma$, and $\rho$ can be
calculated from equation (\ref{D_trho}). 

The background solution is a spatially flat FRW universe
described by $K=0$. The corresponding value of the
energy density (indicated with the suffix ``b'') is
calculated from the Friedmann equation 
\beq 
\lp \frac{\dot s}{s}\rp^2 = H_\mr{b}^2 =
\frac{8\pi}{3}e_\mr{b}\,. \label{Fried_eq} 
\eeq 
The quantity $e_b$ is related to the scale factor $s(t)$
by equation (\ref{M_dot}) used for the unperturbed case
\beq 
\frac{\dot e_b}{e_b} = -3(1+\gamma)\frac{\dot s}{s}
\label{homo_e}\, , 
\eeq
while the  background values of the other quantities are
obtained from the FRW metric (\ref{FRW}) and the set of
equations (\ref{R_dot}) - (\ref{R_prime}): 
\bea
& a_\mr{b} \ug 1 \\
& b_\mr{b} \ug s(t) \\
& R_\mr{b} \ug s(t)r \label{R_b}\\
& M_\mr{b} \ug \frac{4}{3}\pi e_\mr{b}R_\mr{b}^3 \\
& U_\mr{b} \ug H_\mr{b}R_\mr{b} = \dot{s}(t)\,r 
\eea

As already mentioned, when the perturbation is well
outside the cosmological horizon, all of the
hydrodynamical quantities have nearly homogeneous
profiles with their perturbations being small deviations
away from the uniform solution. It is useful therefore to
parametrise the scale of the perturbations using a
dimensionless parameter $\epsilon$ that gives explicitly
the ratio of the cosmological horizon scale $R_H =
H_\mr{b}^{-1}$ to the physical length scale of the
perturbation $R_0 = s(t)r_0$. 
\beq 
\epsilon \equiv \lp \frac{R_H}{R_0}\rp^2 \ug \lp
\frac{1}{H_\mr{b}sr_0}\rp^2 = \frac{1}{\dot s^2
r_0^2}\,<<\, 1 \label{eps_def} 
\eeq 
From this equation it is clear that $\epsilon$ is a
function of time only, and its time derivative, as
follows from (\ref{homo_e}), is given by 
\beq 
\frac{\dot\epsilon}{\epsilon} =
-2\frac{\dot s}{s} - \frac{\dot e_\mr{b}}{e_\mr{b}} =
-2\frac{\dot s}{s} + 3(1+\gamma)\frac{\dot s}{s} =
(1+3\gamma)\frac{\dot s}{s}\,, \label{dot_eps} 
\eeq 
It is important to point out that expression
(\ref{dot_eps}) is a consequence of the energy
conservation law and holds also when $\gamma$ is a
function of time. 

We treat $\epsilon$ as a small parameter for our
asymptotic solution and we will see that it cancels out
in the final equations of this section. However, in
numerical computations we use an explicit value of
$\epsilon$, which for self consistency should be
sufficiently small. In this case we can keep only first
order terms with respect to $\epsilon$ in the
hydrodynamical equations (\ref{R_dot}) -
(\ref{R_prime}). Then the initial perturbations are
defined as:  
\bea
& R = R_b(1+\epsilon\tilde R) \label{R_def} \\
& U = H_bR(1+\epsilon\tilde U) \label{U_def}\\
& b = \frac{R^\pr}{\sqrt{1-K(r)r^2}}(1+\epsilon\tilde
b) \label{b_def}\\
& a = 1+\epsilon\tilde a \label{a_def}\\
& e = e_b(1+\epsilon\tilde e) \\
& M = \frac{4}{3}\pi e_bR^3(1+\epsilon\tilde M) 
\eea 
The tilde quantities are in general functions of both $r$
and $t$. It is convenient to define a new independent
time variable $\xi \equiv \textrm{ln}(s)$ given by
\bea
\frac{\dot{s}}{s}\frac{\partial}{\partial t}  \ug
\frac{\partial}{\partial\xi}\,. \label{xi} 
\eea 
\vspace{2mm}

\noindent 
Starting from equation (\ref{R_dot}), and using
expressions (\ref{R_def}), (\ref{U_def}) and
(\ref{a_def}) we get 
\[ \frac{\dot s}{s} + (\epsilon\tilde R)^{\bdot} =
\frac{\dot s}{s} (1+\epsilon\tilde a) (1+\epsilon\tilde U) \]\,,
then  linearising this equation and using (\ref{dot_eps})
and (\ref{xi}) we have 
\beq
\displaystyle{(1+3\gamma)\tilde R +
\frac{\partial\tilde R}{\partial\xi} = \tilde a + \tilde
U}\label{tilda_R}\,. 
\eeq 
\vspace{2mm}

\noindent 
Similarly, perturbing equation (\ref{b_dot}) we get
\[ (\epsilon\tilde b)^{\bdot} = -\epsilon\frac{\tilde{a}^\pr
U}{R^\pr} \] 
and then, using (\ref{dot_eps}), we have 
\beq
\displaystyle{(1+3\gamma)\tilde b + \frac{\partial\tilde
b}{\partial\xi} = -r{\tilde a}^\pr}\,. \label{tilda_b} 
\eeq
\vspace{2mm}

\noindent 
Perturbing equation (\ref{a_prime}) and integrating it
over $r$ we have 
\beq 
\displaystyle{(1+\gamma){\tilde a} +
\gamma{\tilde e} = 0}\,. \label{tilda_a} 
\eeq 
\vspace{2mm}

\noindent 
Perturbing equation (\ref{M_dot}) to the first
order in $\epsilon$ we get 
\beq
\frac{1}{3}(1+\epsilon\tilde M) \lP \frac{\dot
e_b}{e_b}+3\frac{\dot R}{R}+ (\epsilon\tilde M)^{\bdot}\rP = -
\gamma(1+\epsilon\tilde
e)\frac{\dot R}{R}\,,  
\eeq
and, using (\ref{homo_e}), (42) can be re-written as 
\bea 
& - 3(1+\gamma)\frac{\dot s}{s} + 3(1+\gamma)\frac{\dot R}{R} -
3(1+\gamma)\frac{\dot s}{s}\epsilon\tilde M +
(\epsilon\tilde M)^{\bdot} + \nonumber \\
& \hspace{4.2cm} + 3\frac{\dot R}{R}\epsilon\tilde M +
3\gamma\frac{\dot R}{R}\epsilon\tilde e = 0 \nonumber\,. 
\eea
Using equation (\ref{R_b}) we get
\[
\dot\epsilon\tilde M + \epsilon\dot{\tilde M} - 3\gamma\frac{\dot
s}{s}\epsilon\tilde M + 3\gamma\frac{\dot s}{s}\epsilon\tilde e +
3(1+\gamma)(\dot\epsilon\tilde R + \epsilon\dot{\tilde R}) = 0 \,,
\]
and finally using equations (\ref{dot_eps}), (\ref{xi}),
(\ref{tilda_R}) and (\ref{tilda_a}) we have 
\beq
\displaystyle{\tilde M + \frac{\partial\tilde M}{\partial\xi} =
-3(1+\gamma)\tilde U}\,. \label{tilda_M} 
\eeq 
\vspace{2mm}

\noindent 
Perturbing equation (\ref{M_prime}) we obtain
\[
(1+\epsilon\tilde M) \frac{R^\pr}{R} + \frac{1}{3}\epsilon{\tilde
M}^\pr = (1+\epsilon\tilde e)\frac{R^\pr}{R}\,,
\]
that can be re-written as \beq \displaystyle{\tilde e =
\frac{1}{3r^2}(r^3 \tilde{M})^\pr}\,. \label{tilda_e} \eeq
\vspace{2mm}

\noindent 
Perturbing equation (\ref{R_prime}), we have
\[
\lP (1-K(r)r^2)(1-2\epsilon\tilde b) - 1\rP = \frac{\dot
{s^2}}{s^2}R^2(1+2\epsilon\tilde U) - \frac{8\pi
e_\mr{b}R^2}{3}(1+\epsilon\tilde M)\,,
\]
that using (\ref{Fried_eq}) becomes
\[
- \lP K(r)r^2 + 2\epsilon\tilde b(1-K(r)r^2)\rP = R^2\frac{\dot
s^2}{s^2}\epsilon(2\tilde U - \tilde M)\,,
\]
and now using expression (\ref{eps_def}) for $\epsilon$ we have
\beq 
- \lP K(r)r^2 + 2\epsilon\tilde b(1-K(r)r^2)\rP =
\frac{r^2}{r_0^2}(2\tilde U - \tilde M)\,. \nonumber
\label{K_Gamma} 
\eeq 
Since $\epsilon\ll1$, one can drop the last term on
left-hand side, obtaining finally  
\beq
\displaystyle{\tilde U= \frac{1}{2}\lP \tilde M - K(r)r_0^2\rP}\,.
\label{tilda_U} 
\eeq 
This is the only equation of the system where $K(r)$
appears explicitly.  From the definition of $\Gamma$
given by (\ref{Gamma}) one can rewrite (\ref{tilda_U}) as
\beq 
K(r) \ug \frac{1-\Gamma^2}{r^2}\,. \label{K_Gamma2}
\eeq 
This relation shows also the connection between the
profile $K(r)$ and $\Gamma$, which is an invariantly
defined quantity representing an energy per unit mass.

Substituting (\ref{tilda_U}) into equation (\ref{tilda_M}) we get
\beq 
\frac{\partial\tilde M}{\partial\xi} +
\frac{5+3\gamma}{2}\tilde M = \frac{3}{2}(1+\gamma)K(r)r_0^2\,,
\eeq 
and from this expression we can see that it is possible
to separate the variables $(r,\xi)$ because, as we have
assumed at the beginning, $\gamma$ is just a function of
time. We therefore write 
\beq \displaystyle{\tilde M = \Phi(\xi)K(r)r_0^2}\,,
\label{M_tilda} 
\eeq 
that gives the following differential equation for the
function $\Phi(\xi)$ 
\beq
\displaystyle{\frac{d\Phi}{d\xi} + \frac{5+3\gamma}{2}\Phi =
\frac{3}{2}(1+\gamma)}\,. \label{Phi_equation} 
\eeq 
Expression (\ref{M_tilda}) is the solution for the mass
perturbation $\tilde M$ and inserting this into
(\ref{tilda_e}), (\ref{tilda_a}) and (\ref{tilda_U}) we
obtain the following expressions for first order
perturbations of energy density, $\tilde e$, lapse,
$\tilde a$ and radial velocity, $\tilde U$: 
\bea 
& \tilde e = \Phi(\xi)
\frac{1}{3r^2}\lP r^3K(r)\rP^\pr r_0^2
\label{e_tilda}\\
& \tilde a =  - \Phi(\xi)\frac{\gamma}{1+\gamma} \frac{1}{3r^2}\lP
r^3K(r)\rP^\pr r_0^2
\label{a_tilda} \\
& \tilde U = \frac{1}{2}\lP \Phi(\xi)-1\rP K(r)r_0^2
\label{U_tilda} 
\eea 
Substituting these equations into (\ref{tilda_R}) and
(\ref{tilda_b}) we find two differential equations for
$\tilde R$ and $\tilde b$, given by 
\bea 
& (1+3\gamma)\tilde R + \frac{\partial\tilde
R}{\partial\xi} = - \Phi(\xi)\frac{\gamma}{1+\gamma}
\frac{1}{3r^2}\lP r^3K(r)\rP^\pr r_0^2 \nonumber \\
& \hspace{5cm} +\frac{1}{2}\lP \Phi(\xi)-1\rP K(r)r_0^2\,,
\label{R_tilda_eq} \\
& (1+3\gamma)\tilde b + \frac{\partial\tilde b}{\partial\xi} =
\frac{\gamma}{(1+\gamma)}r \lP \frac{1}{3r^2}\lP
r^3K(r)\rp^\pr\rP^\pr \Phi(\xi)\,, \label{b_tilda_eq} 
\eea  
that can be solved introducing two new functions of time
$I_1(\xi)$ and $I_2(\xi)$ and writing 
\bea 
& \tilde R = - I_1(\xi) \frac{1}{3r^2}\lP r^3K(r)\rP^\pr
r_0^2 + I_2(\xi)\frac{K(r)}{2} r_0^2 \,, \label{R_tilda} \\
& \tilde b = I_1(\xi)r\lP \frac{1}{3r^2}\lp r^3K(r)\rp^\pr\rP^\pr
r_0^2\,. \label{b_tilda} 
\eea 
The relations between $I_{1,2}(\xi)$ and $\Phi(\xi)$ are
obtained by introducing (\ref{R_tilda}) and
(\ref{b_tilda}) into (\ref{R_tilda_eq}) and
(\ref{b_tilda_eq}): 
\bea 
& \frac{dI_1(\xi)}{d\xi} + (1+3\gamma)I_1(\xi) =
\frac{\gamma}{1+\gamma}\Phi(\xi) \label{I1} \\
& \frac{dI_2(\xi)}{d\xi} + (1+3\gamma)I_2(\xi) = [\Phi(\xi)-1]
\label{I2} 
\eea 
This completes the system of equations.
It worth mentioning that in these equations, time
evolution is given by $\epsilon$ and $\Phi$, while all
spatial dependencies are given by $K(r)$.

\section{Perturbations in a multifluid medium}
In general the universe consists of different fluid
components characterised by different constant values
$\gamma_i$. We can introduce an effective $\gamma$
defined as 
\beq \gamma(s)
\equiv \frac{\sum_i p_i}{\sum_i e_i} = \frac{\sum_i \gamma_i
e_i}{\sum_i e_i} = \frac{\sum_i \gamma_i e_i}{e}\,,
\label{gamma_def} 
\eeq 
which, in general, is a function of time. At some moment
of time $t_0$ (which is not necessarily the same moment 
when we impose the initial conditions) we specify the
fractional contributions $f_i$ of the different fluid
components with the corresponding coefficients
$\gamma_i$, having for each component  
\bea 
& \frac{e_i}{e_{i0}} \ug \lp \frac{s}{s_0}
\rp^{-3(1+\gamma_i)} \label{e_i1} \\
& e_i = f_i\,e \label{e_i2} \\
& \sum_i f_i= 1 
\eea 
Using expressions (\ref{e_i1}) and (\ref{e_i2}) one can
write 
\beq 
e_i \ug f_{i0}\,e_{0} \lp
\frac{s}{s_0} \rp^{-3(1+\gamma)} 
\eeq 
then from (\ref{gamma_def}) and the equation of state
(\ref{eq.state}) we have  
\beq 
\gamma = \frac{\sum_i f_{i0}\gamma_i
s^{-3(1+\gamma_i)}}{\sum_i f_{i0} s^{-3(1+\gamma_i)}}. 
\eeq 
We now define a new quantity $Q$ 
\beq
Q(s) \equiv \sum_i f_{i0} s^{-3(1+\gamma_i)} \propto
e(s)\,. 
\eeq
Taking the time derivative of this quantity, one gets
\beq
\frac{dQ}{d\xi} = s\frac{dQ}{ds} =  - 3\sum_i
(1+\gamma_i)f_{i0}\ s^{-3(1+\gamma_i)} = - 3(1 + \gamma)Q
\label{dQ/dxi}
\eeq
and finally $Q$ can be related to $H_b$ by the following
differential equation: 
\beq 
\frac{1}{2}\frac{dQ}{Qd\xi} =
\frac{1}{H_b}\frac{dH_b}{d\xi}\,. \label{gamma} 
\eeq 
Using expressions (\ref{dQ/dxi}) and (\ref{gamma}) into
(\ref{Phi_equation}) one obtains 
\beq
\frac{d\Phi}{d\xi} + \lp 1-\frac{1}{H_b}\frac{dH_b}{d\xi}\rp\Phi =
- \frac{1}{H_b}\frac{dH_b}{d\xi}\,.
\eeq
The solution of this equation can be written in the form
\[
\Phi = Fe^{\lambda}\,,
\]
where $F$ and $\mu$ satisfy the following differential equations
\[
\frac{dF}{d\xi} = -\frac{1}{H_b}\frac{dH_b}{d\xi}e^{-\lambda}\,,
\]
\[
\frac{d\lambda}{d\xi} + 1 -\frac{1}{H_b}\frac{dH_b}{d\xi} = 0 \,,
\]
which give
\[
e^{-\lambda} = e^{\textrm{ln}s-\textrm{ln}H_b} = \frac{s}{H_b} \,,
\]
\[
F = \frac{s}{H_b} - \int_0^s \frac{ds}{H_b}\,,
\]
and finally 
\beq 
\displaystyle{\Phi(s) = 1 - \frac{H_b(s)}{s}
\int_0^s \frac{ds}{H_b(s)}}\,. 
\eeq

Now we can solve the two differential equations for $I_1$ and
$I_2$ that are both of the same form, 
\beq 
s\frac{dI_i(s)}{ds} + (1+3\gamma)I_i(s) = F_i(s)\,,\quad
i=1,2 \label{I_eq}  
\eeq 
where $F_1(s) = \frac{\gamma(s)}{1+\gamma(s)}\Phi(s)$ and
$F_2(s) = \Phi(s) - 1$. From the expression
(\ref{dot_eps}) we see that 
\beq
1+3\gamma = \frac{1}{H_b\epsilon}\frac{d\epsilon}{dt} =
H_b^2s^2\lp s\frac{d(1/H_b^2s^2)}{ds}\rp \nonumber 
\eeq 
and this allows the differential equations (\ref{I_eq})
to be written in a form that can easily be integrated, 
\beq 
s\frac{d}{ds}\lP
\frac{1}{H_b^2s^2}I(s)\rP = \frac{1}{H_b^2s^2}F(s) 
\eeq 
therefore
\beq 
I(s) = H_b^2(s)\,s^2 \int_0^s
F(s)\frac{1}{H_b^2(s)\,s^2}\frac{ds}{s}\,, 
\eeq 
which gives the following expressions for the two
functions $I_1$ and $I_2$: 
\beq
\displaystyle{I_1(s) = H_b^2(s)\,s^2 \int_0^s
\frac{\gamma(s)}{1+\gamma(s)}\Phi(s)\frac{1}{H_b^2(s)\,s^2}\frac{ds}{s}}\,,
\eeq 
\beq \displaystyle{I_2(s) = H_b^2(s)\,s^2 \int_0^s
(\Phi(s)-1)\frac{1}{H_b^2(s)\,s^2}\frac{ds}{s}}\,. 
\eeq

The initial conditions for all physical quantities are
now determined in a self consistent way: the functions
$Q(s)$ and $H_b(s)$ specify the multifluid medium
consisting of different species of particles and fields
(for example massive particles, photons, scalar fields,
cosmic strings, etc.) which are present in the Universe,
and hence in the configuration, at any time preceding the
time chosen for imposing the initial conditions. In the
particular case of a single fluid component, with
$\gamma\,=\, const$, the functions $\Phi$, $I_1$, and
$I_2$ are also constants. We have in this case 
\beq
H(s)\,\propto\,s^{-\frac{3(1+\gamma)}{2}}, 
\eeq 
and 
\bea
& \Phi = \frac{3(1+\gamma)}{5+3\gamma} \\
& I_1 = \frac{\gamma}{(1+3\gamma)(1+\gamma)}\Phi =
\frac{3\gamma}{(1+3\gamma)(5+3\gamma)} \\
& I_2 = \frac{1}{1+3\gamma}(\Phi - 1) =
-\frac{2}{(1+3\gamma)(5+3\gamma)} 
\eea 
Therefore in this simple case the tilde-quantities are
time independent and the time evolution is determined
only by the parameter $\epsilon$. From (\ref{dot_eps}) we
can calculate the time evolution of $\epsilon$ 
\beq 
\epsilon(t) \propto
\lp\frac{t}{t_0}\rp^{\frac{2(1+3\gamma)}{3(1+\gamma)}}\,,
\label{lin_regime} 
\eeq 
which is the same as the standard solution for a growing
mode in cosmological perturbation theory (see for
example \cite{Liddle,Padmanabhan}).

\section{General properties of the curvature profile}

\subsection{The curvature profile and the comoving curvature perturbation} 
The curvature profile $K(r)$ used in the present paper is
linked to the comoving curvature perturbation
$\mathcal{R}$ frequently used in literature (see for
example, \cite{Liddle}). The simplest definition of
$\mathcal{R}$ comes from the theory of linear
perturbations in the FRW metric, where the spatial part
of the metric tensor is given by  
\beq 
g_{ij} \ug
s^2\lP\lp1+2\mathcal{R}\rp\delta_{ij}\rP\,. 
\label{g_ij_linear}
\eeq 
The calculation of the spatial scalar curvature $R^{(3)}$
gives 
\beq 
R^{(3)} \ug 4\frac{k^2}{s^2}\mathcal{R}\,, 
\label{R3_1}
\eeq 
where $k$ is the comoving wavenumber associated with the
perturbation. It is possible to generalise the definition
of $\mR$ for non linear perturbations of the metric
\cite{Lyth_et_al,Lan_Ver}, with $\mR$ being still time
independent (as first shown in 1963 by Lifshitz \&
Khalatnikov \cite{Lifshits}) and converging to
(\ref{g_ij_linear}) for small perturbations. Starting
from the expression for the scalar curvature given by
(\ref{R3_1}) for generic metric perturbations one can see
that the comoving curvature perturbation $\mathcal{R}$ is
connected with the pure growing mode in the case of small
density perturbations
\beq 
\frac{\delta e}{e_\mr{b}} \ug \frac{2(1+\gamma)}{5+3\gamma}
\lp\frac{k}{sH}\rp^2 \mathcal{R} \,, 
\label{comoving_curv} 
\eeq
where $\gamma$ is constant. This relation is obtained
using a quantity called the ``peculiar gravitational
potential'' $\Phi_\mr{G}$ defined as 
\beq 
\frac{\delta e}{e_\mr{b}} \ug - \frac{2}{3}
\lp\frac{k}{sH}\rp^2 \Phi_\mr{G}\,, 
\label{Phi_G1} 
\eeq 
and linked to $\mathcal{R}$ by a first order differential
equation 
\beq 
\frac{1}{H}\dot{\Phi}_\mr{G} +
\frac{5+3\gamma}{2}\Phi_\mr{G} \ug
-(1+\gamma)\mathcal{R}\,. 
\label{Phi_G} 
\eeq 
This equation has a pure growing solution given by 
\beq 
\Phi_\mr{G} \ug
-\frac{3(1+\gamma)}{5+3\gamma}\mathcal{R}\,, 
\eeq 
where $\mathcal{R}$ and $\Phi_\mr{G}$ are time
independent when the length-scale of a perturbation is
much larger than the horizon scale. If we calculate the
spatial curvature scalar (see appendix) in terms of the
curvature profile $K(r)$, we find that 
\beq 
R \ug 6\lP \frac{\ddot{s}}{s} \lp 1 + \epsilon
\frac{3\gamma-1}{5+3\gamma}\mathcal{K} \rp + \frac{\dot{s}^2}{s^2}
\lp 1 - \epsilon \frac{2(2+3\gamma)}{5+3\gamma}\mathcal{K} \rp +
\frac{\mathcal{K}}{s^2} \rP\,, \label{R4} 
\eeq 
where 
\beq 
\mathcal{K} \ug K(r) + \frac{r}{3}K^\prime(r)\,. 
\eeq 
For the diagonal metric (\ref{sph_metric}) the three
curvature is given by 
\beq 
R^{(3)} \ug \frac{6}{s^2} \lP K(r)+\frac{r}{3}K^\pr(r) \rP\,, 
\label{R3_2} 
\eeq 
because only the last term in the square brackets of
equation (\ref{R4}) corresponds to the diagonal spatial
component of the Ricci tensor. We now identify 
\beq 
k \ug 1/r_0\,, 
\label{k_r0} 
\eeq 
which is obvious to do because the wave number represents the
inverse length of the perturbation. Comparing  equations
(\ref{R3_1}) and (\ref{R3_2}), one can then see that the two
equations are the same if 
\beq 
\mathcal{R} \ug \frac{3}{2}r_0^2 \lP K(r)+\frac{r}{3}K^\pr(r) \rP\,.
\label{cuvR_curvK} 
\eeq 
From (\ref{k_r0}) and (\ref{eps_def}) it follows also
that 
\beq \epsilon \ug \lp\frac{k}{sH}\rp^2\,
\label{eps_k} 
\eeq 
and relation (\ref{cuvR_curvK}) is obtained again by
comparison of (\ref{e_tilda}) with
(\ref{comoving_curv}). We can notice also that equation
(\ref{Phi_G})  is the same as (\ref{Phi_equation}) since
the peculiar gravitational potential defined in
(\ref{Phi_G1}) can be written in our terminology as
$\Phi_\mr{G} \ug -(2/3)\tilde{e}$ and we can then use
equation (\ref{tilda_e}) to express the variable
$\tilde{M}$ in terms of $\tilde{e}$ in
(\ref{Phi_equation}). We have therefore shown that
$\mathcal{R}$ is directly linked to $K(r)$. 

\subsection{Physical properties of the curvature profile}
From the definition of $b$ given by (\ref{b_def}) we obtain
obvious mathematical requirement 
\beq 
1 - K(r)r^2 > 0 \ \ \Rightarrow \ \ K(r) < \displaystyle{\frac{1}{r^2}}\,.
\label{cond1} 
\eeq 
This corresponds to the physical condition that a
perturbed spherical region of comoving radius $r$ should
not be causally disconnected from the rest of the
Universe. Another important requirement is related to
conservation of total mass, given by 
\beq 
\int_0^\infty 4\pi r^2\tilde{e}(r)dr = 0\,,
\label{cond2_0} 
\eeq 
that implies 
\beq 
\lim_{r \rightarrow \infty} r^3K(r) = 0\,. 
\label{cond2} 
\eeq 
This means that the Universe remains spatially flat at
infinity. As explained in \cite{Nadezhin}, if condition
(\cite{cond2_0}) is not satisfied one unavoidably
confronts a violation of the causality principle. 

An important parameter characterising the curvature
profile is the value of $r_0$ that specifies the comoving
length-scale of the overdense region in the
configuration. From (\ref{e_tilda}) we have 
\beq 
\tilde{e}(r_0) = 0 \ \ \Rightarrow \ K(r_0) +
\displaystyle{\frac{r_0}{3}K^\pr(r_0) \ug 0}\,. 
\label{cond3} 
\eeq
It is useful to introduce an integrated quantity $\delta$
that measures the mass-excess inside the overdense
region, as frequently done in literature. From equation
(\ref{e_tilda}), one can see that any spherical integral
of $\tilde{e}$ does not depend on the particular
curvature profile used, but depends only on the value of
$K(r)$ at the outer edge of the configuration. The
expression for $\delta$ is then given by  
\beq 
\delta \equiv \lp
\frac{4}{3}\pi r_0^3\rp^{-1} \displaystyle{\int_0^{r_0} 4\pi
\frac{e-e_b}{e_b}r^2dr} \ug \epsilon(s)\Phi(s)K(r_0)r_0^2\,.
\label{delta_s} 
\eeq 
We can see from this expression that $\delta$ is a gauge
invariant quantity since it is directly proportional to
$K(r_0)r_0^2$, which is also a gauge invariant
quantity. Note once more the perfect separation of the
different physical aspects of the problem: $\epsilon$
specifies the evolution of the perturbation amplitude in
the long wavelength limit, $\Phi$  represents the
equation of state and $K(r)r^2$ specifies the spatial
profile of the configuration. 

\section{Parametrisation of the curvature profile}
In this section we specify the profile  $K(r)$ in terms
of two free parameters. First we want to point out that
both $r$ and $K$ are gauge-dependent
quantities. Therefore we need to be careful when
transferring the initial perturbation profiles onto the
grid used for the numerical calculations. The
circumference coordinate $R_b$ used in the code is gauge
invariant and is related to $r$ by $R_b = s(t)r$, while
particular values of $s(t)$ and $r$ are gauge
dependent. Another gauge invariant quantity is the
product $K(r)r^2$. These two gauge invariant products are
connected by the presence of the comoving coordinate in
both of them and therefore a particular specification of
one of these three variables will specify the others.

To do this we compare the length scale $R_{0}$ of the
configuration with the cosmological horizon $R_H$ 
\beq 
R_{0} \ug N R_H\,, 
\label{N} 
\eeq 
where $N$ is the number of horizon scales inside the
configuration. From the definition of $\epsilon$ in
(\ref{eps_def}), we have 
\beq 
\epsilon \ug \frac{1}{N^2}\,. 
\eeq
To have $\epsilon\,\ll\,1$ we need to have $N\,\gg\,1$
and $N$ will be one of the input parameters used to
impose the initial conditions. Expressions (\ref{N}) and
(\ref{eps_def}) relate the scale factor with  $r_0$ \beq
s(t) \ug \frac{NR_H}{r_0}\,, \label{sc_fac} \eeq and we
can use this to define the comoving coordinate as 
\beq 
r \ug \frac{R_\mr{b}}{s(t)} \ug
\frac{r_0}{NR_H}R_\mr{b}\,, 
\label{r_def} 
\eeq 
where the value of $R_\mr{b}$ is calculated from equation
(\ref{d_mu})  
\beq 
R_\mr{b} \ug \lp\frac{\mu}{4\pi\rho}\rp^{1/3}\,, 
\eeq 
and $r_0$ is determined by condition (\ref{cond3}) once
the curvature profile is specified.

To parametrise $K(r)$ we should decide what is a
reasonable perturbation shape. It is natural to start
with a centrally peaked profile of the energy density,
such that outside the overdense region there is an
under-dense region which becomes asymptotically flat. To
obtain an energy density profile with these properties,
one should choose a centrally peaked curvature profile
which is described in terms of a suitable continuous
function and tends asymptotically to zero as
$r\rightarrow\infty$. Continuity should be ensured at
least in the first and second derivatives, because these
derivatives play a crucial role in the expressions for
the perturbation profiles of density and velocity. To
preserve the standard normalisation used in cosmology
where a closed universe has a curvature equal to $1$, we
choose to normalise the curvature profile to have
$K(0)=1$. 

We start with a family of curvature profiles based on a
Gaussian shape, described as 
\beq 
K(r) = \lp 1+\alpha\frac{r^2}{\Delta^2}\rp \exp{\lp
-\frac{r^2}{2\Delta^2}\rp} 
\label{curv_profile} 
\eeq 
where $\alpha$ and $\Delta$ are two independent
parameters. In the particular case $\alpha = 0$, the
profile $K(r)$ is exactly Gaussian. The first derivative
of $K(r)$  is given by 
\beq 
K^\pr(r) = \frac{r}{\Delta^2}\lP \alpha - \lp
1+\alpha \frac{r^2}{\Delta^2}\rp \rP \exp{\lp
-\frac{r^2}{2\Delta^2}\rp}\,, 
\label{curv_dev1} 
\eeq 
and we use this to calculate the value of $r_0$ as
function of $\alpha$ and $\Delta$ using
(\ref{curv_profile}) and (\ref{cond3}). We obtain a
quadratic equation with the following solution for $r_0$:
\bea 
& r_0^2 \ug 3\Delta^2 \quad \mr{if}\quad \alpha \ug 0
\label{r0_eq} \\
& \nonumber\\
& r_0^2 \ug \Delta^2 \frac{(5\alpha-2) +
\sqrt{(5\alpha-2)^2+24\alpha}}{2\alpha} \quad
\mr{if}\quad \alpha\ne 0 
\label{r0_sol} 
\eea 
that we substitute into (\ref{r_def}) to express the
spatial comoving coordinate $r$ in terms of the
background coordinate $R_\mr{b}$. It is then convenient
to express  $r_0$ in terms of the initial perturbation
length scale $R_0$. We therefore introduce a
dimensionless ratio of the  circumferential radius to
$R_0$ defined as $Z\equiv R_b/R_0$, which is equal to $1$
at the edge of the overdense region. Thus the exponential
argument in (102) is given by  
\beq
\frac{r^2}{\Delta^2} \ug \frac{F(\alpha)}{2\alpha}Z^2\,, \label{Z}
\eeq 
where 
\bea 
& F(\alpha) \,\equiv\,
(5\alpha-2)+\sqrt{(5\alpha-2)^2+24\alpha} \quad \mr{if} 
\quad \alpha \neq 0  \\
& \nonumber \\
&  \frac{F(\alpha)}{\alpha} \ug 3 \quad \mr{if} \quad \alpha \ug 0
\eea
\begin{figure}[t!]
\centerline{\psfig{file=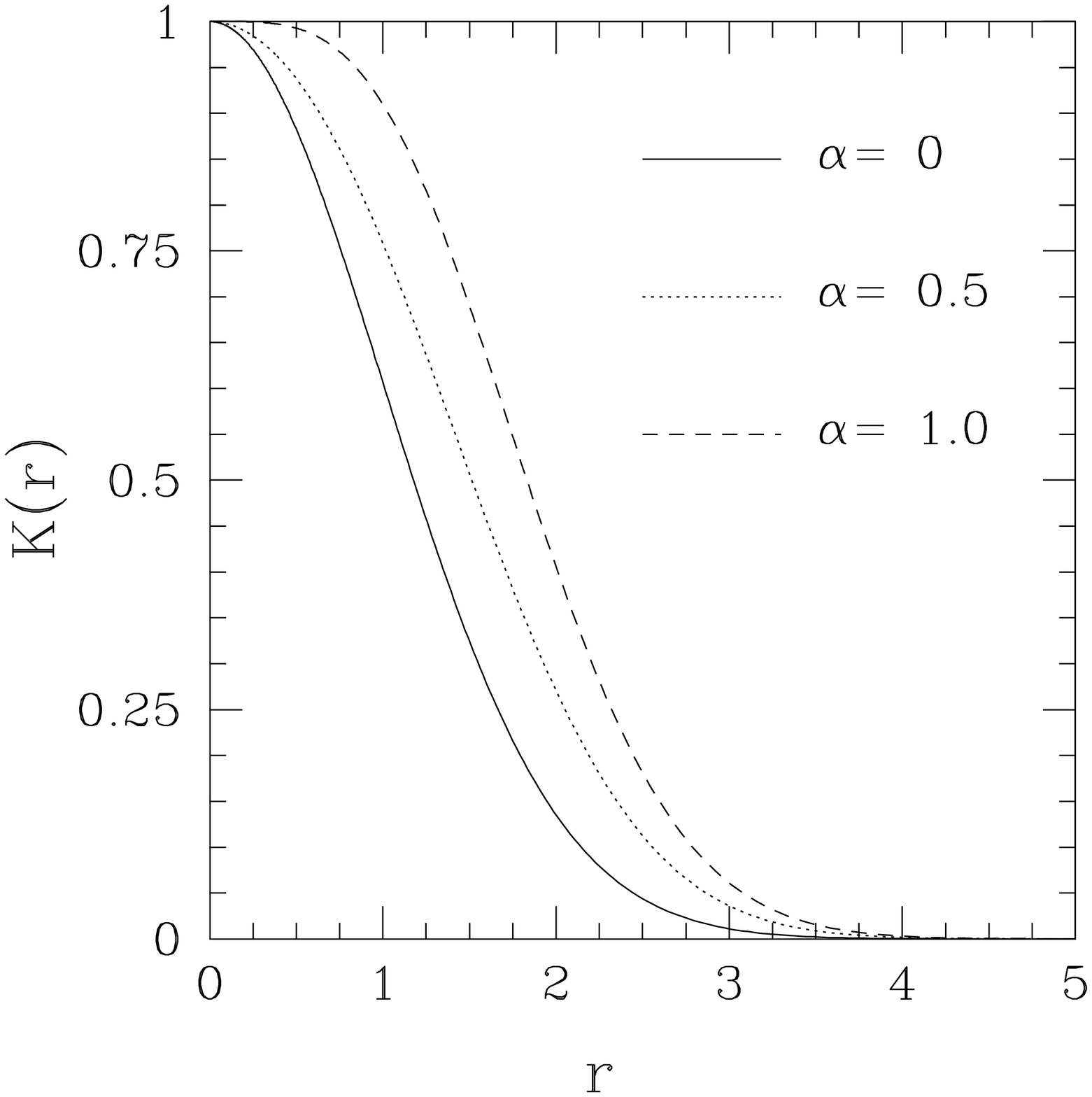,width=6.5cm}
            \psfig{file=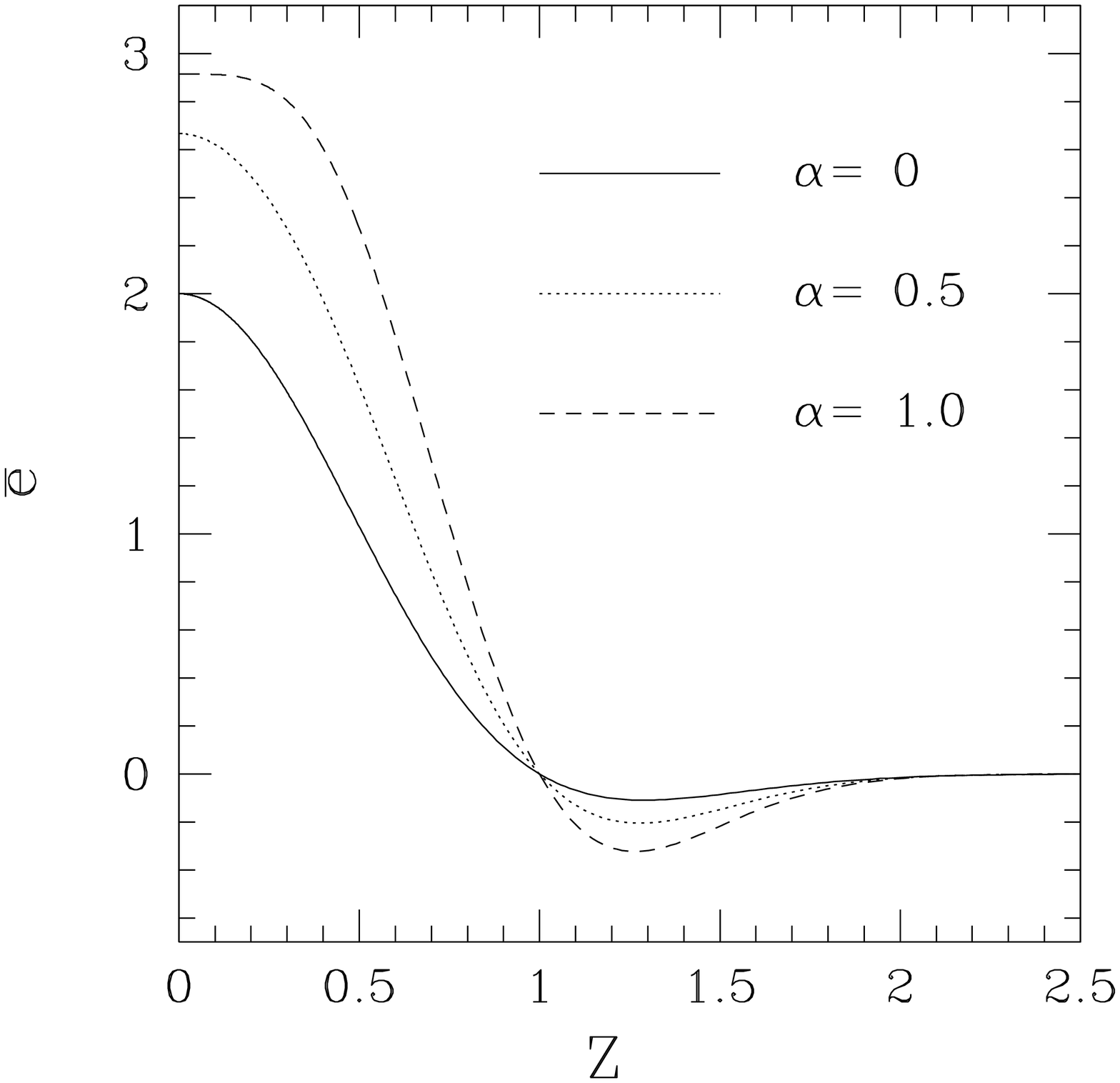,width=6.5cm}}
\caption{\label{profiles1}\small The left hand  plot
shows the curvature profile $K(r)$ as a function of the
comoving coordinate for three different value of $\alpha$
($0$, $0.5$ and $1$). The right hand plot shows the
corresponding profiles of the energy density perturbation
$\tilde{e}$ plotted as functions of $Z$. These cases and
those described in the next figures have been calculated
with $\gamma\ug1/3$ (equivalent to $\Phi\ug2/3$.)}
\end{figure}
and one can see that the parameter $\Delta$ cancels in
the expression for $r/r_0$, making clear the physical
meaning of $\alpha$ and $\Delta$: for the energy density
perturbation profile expressed in terms of  $Z$ we get 
\bea 
& \!\!\!\!\!\!\!\!\!\!\!\!\!\!\!\!\!\!\!\!\!\!\!\!
\tilde{e}(Z)  \ug  \Phi(\xi) \frac{\Delta^2}{2\alpha}
F(\alpha) \lP 1 \,+\, \lp\frac{5}{6}\alpha-1\rp
\frac{F(\alpha)}{2\alpha}Z^2 \,-\, \frac{\alpha}{2}
\lp\frac{F(\alpha)}{2\alpha}\rp^2 Z^4 \rP \times\nonumber\\
& \quad\times\exp{\lp -\frac{F(\alpha)}{4\alpha}Z^2\rp}
\label{e_Z} 
\eea 
and we see that the spatial profile given by the
expression inside the parentheses depends only on
$\alpha$. Hence $\Delta$, appearing only outside the
brackets, parameterises the amplitude of the density
perturbation. The separation of factors in (\ref{e_Z})
allows us to control the values of the input parameters
independently. 

The value of the second derivative of $K(r)$ at the
centre is given by 
\beq 
K^{\pr\pr}(0) \ug \frac{\alpha-1}{\Delta^2}\,. 
\eeq
When $\alpha\,>\,1$, the value $K^{\pr\pr}(0)$ is
positive and an off-centred peak appears in the profiles
of both curvature and energy density. For $\alpha\,<\,0$,
there is a second overdense region, corresponding to a
second solution for $r_0$. Because we are interested in
solutions that represent perturbations centred at $r=0$
surrounded by an underdense region which becomes
asymptotically flat, we restrict our attention to the
range of shapes given by $0\,\leq\,\alpha\,\leq\,1$.

\begin{figure}[t!]
\centerline{\psfig{file=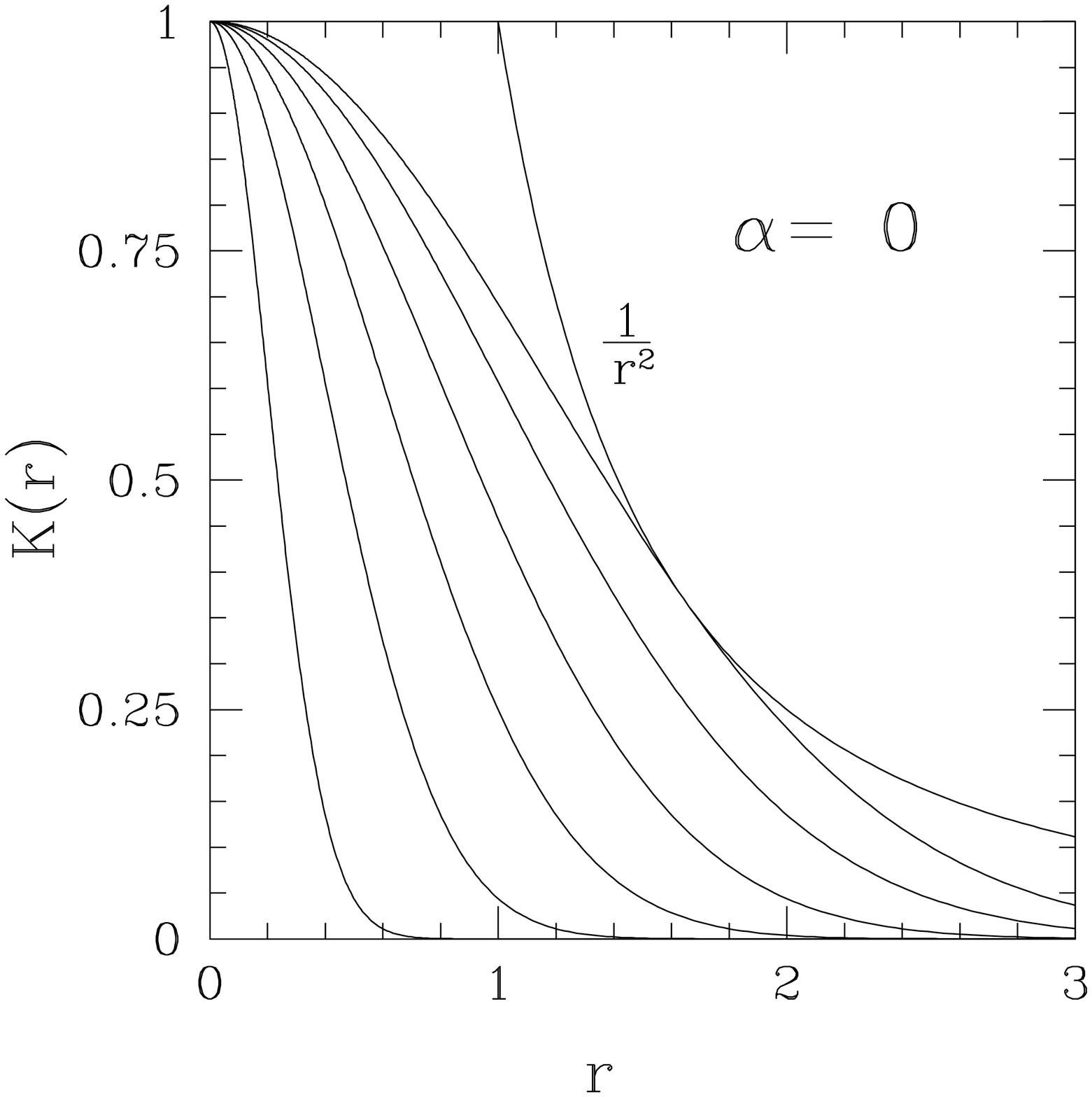,width=6.5cm}
            \psfig{file=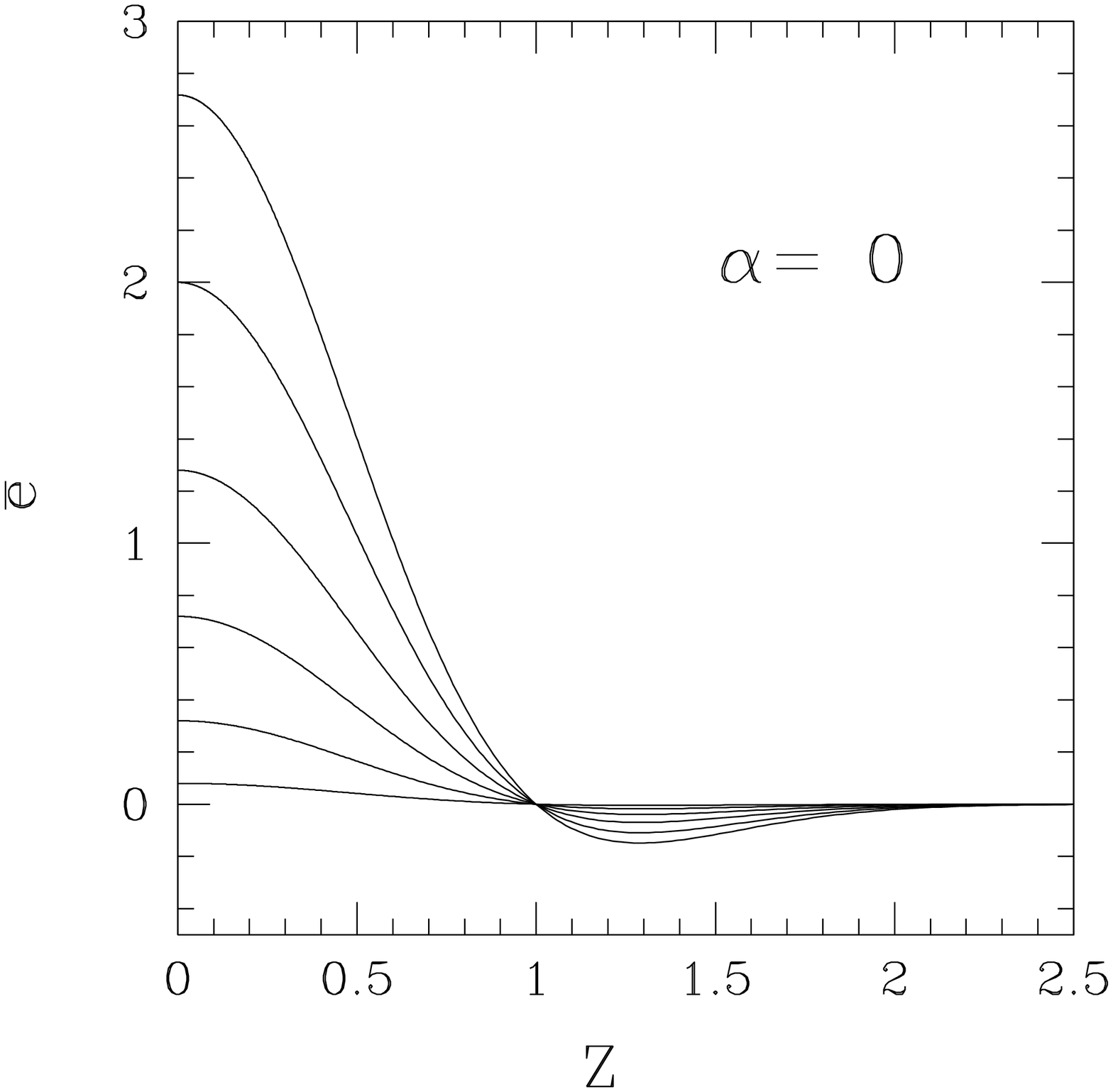,width=6.5cm}}
\centerline{\psfig{file=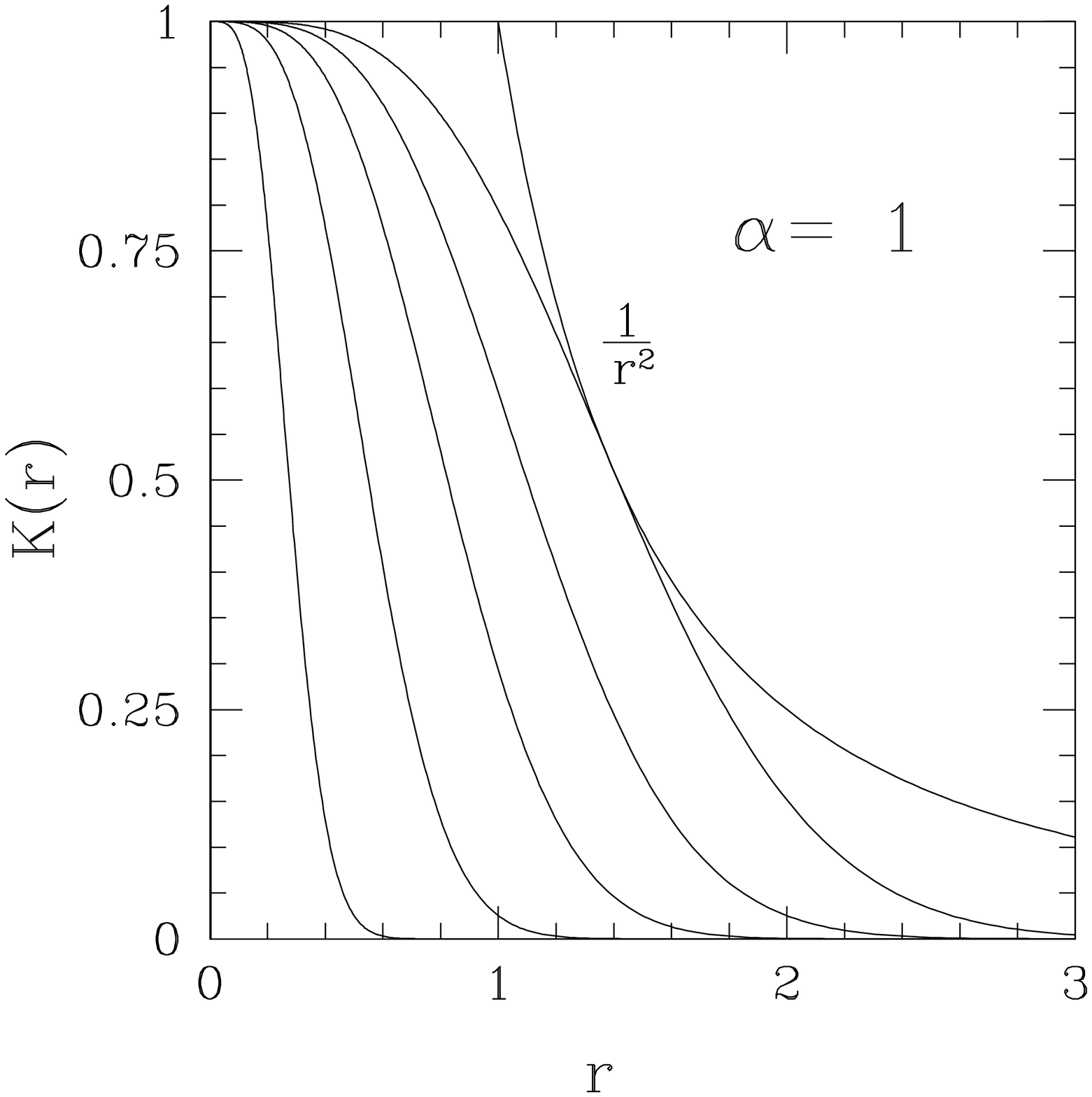,width=6.5cm}
            \psfig{file=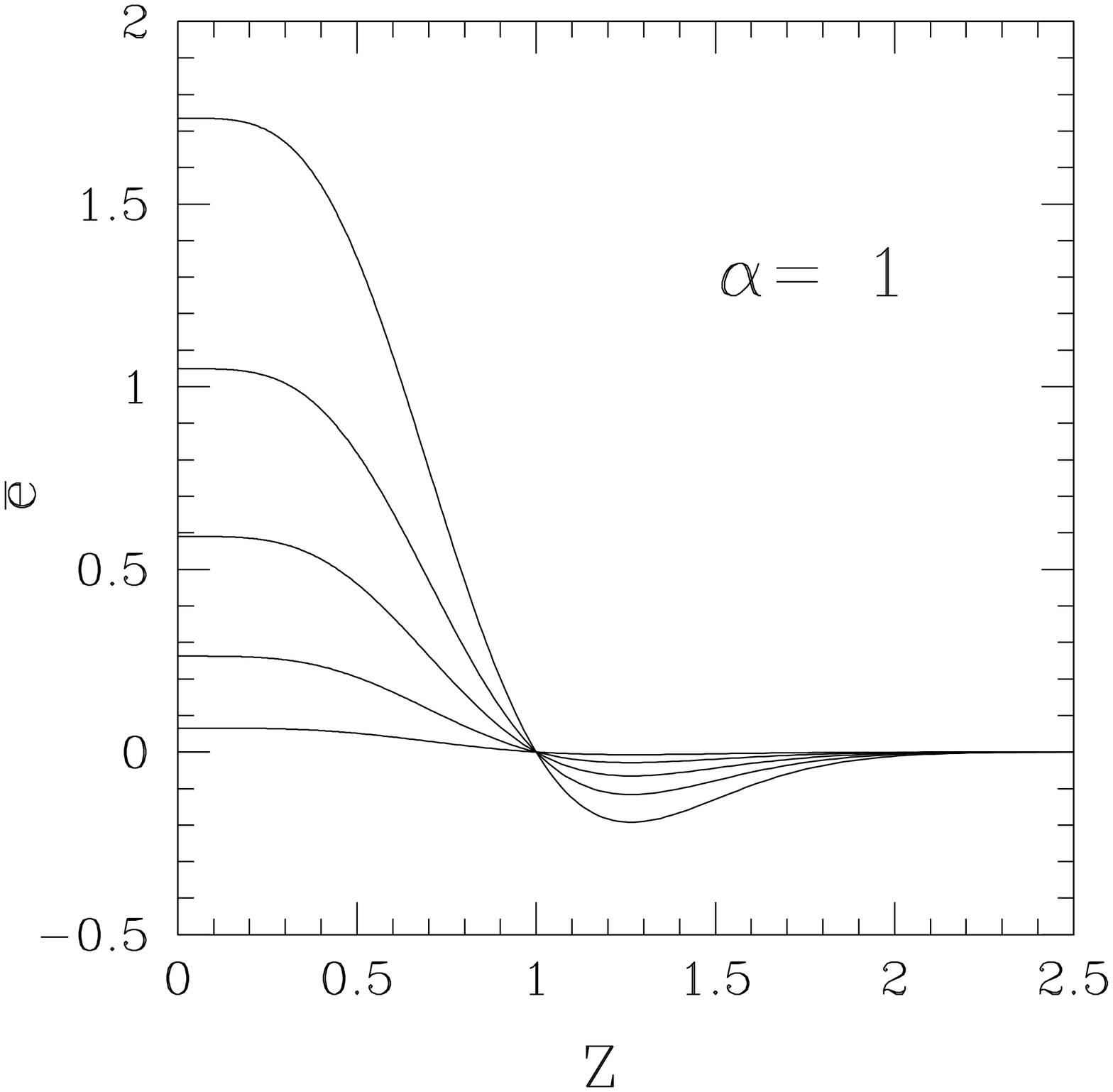,width=6.5cm}}
\caption{\label{profiles2}\small The left hand plots show the
curvature profiles $K(r)$ as functions of the comoving coordinate
$r$, while the right hand plots show the corresponding profiles
for the energy density perturbations $\tilde e$ plotted as
functions of $Z$. For $\alpha\ug0$ (upper plots) the different
profiles correspond to $\Delta$ between $0.2$ and $1.16$, which is
the maximum value allowed by (\ref{cond1}). The values of $\Delta$
are higher for the higher curves. For $\alpha\ug1$ (lower plots),
the values of $\Delta$ used are between $0.15$ and $0.77$.}
\end{figure}

In Figure \ref{profiles1} we plot corresponding curvature and
energy density profiles for three different values of $\alpha$
between $0$ and $1$, keeping the same value of $\Delta$.

The particular case with $\alpha=0$ in (\ref{e_Z})
coincides with one of the profiles used in
\cite{Niemeyer1,Musco}:  
\beq
\tilde{e}(Z)  \ug  \Delta^2 \lP 1 \,+\, \frac{3}{2}Z^2 \rP
\exp{\lp -\frac{3}{2}Z^2\rp}\,, \label{mex_hat} 
\eeq 
where the parameter $\Delta$ has the same meaning as the
quantity used in \cite{Niemeyer1,Musco} to parametrise
the amplitude of the density perturbation.

An advantage of the present description is the
possibility to calculate the perturbation amplitude
$\delta$, defined in (\ref{delta_s}), directly from the
curvature profile. Using (\ref{mex_hat}) for an arbitrary
value of $\alpha$ we have 
\beq 
\delta \ug \frac{1}{3N^2}\frac{\Delta^2}{\alpha}F(\alpha) \lp
1+\frac{F(\alpha)}{4} \rp \exp\lp -\frac{F(\alpha)}{4\alpha}\rp\,.
\label{delta_par} 
\eeq 
Another important point is given by condition
(\ref{cond1}), which should be taken into account to
calculate the maximum allowed value of the density
perturbation amplitude for each profile. Therefore one
needs to calculate the maximum value of $\Delta$ for each
$\alpha$. The results of these calculations are presented
in Figure \ref{profiles2} for two different cases
($\alpha\ug0$ and $\alpha\ug1$). From these plots, we can
see clearly how the density perturbation amplitude
$\delta$ varies with $\Delta$.

\begin{figure}[t!]
\centerline{\psfig{file=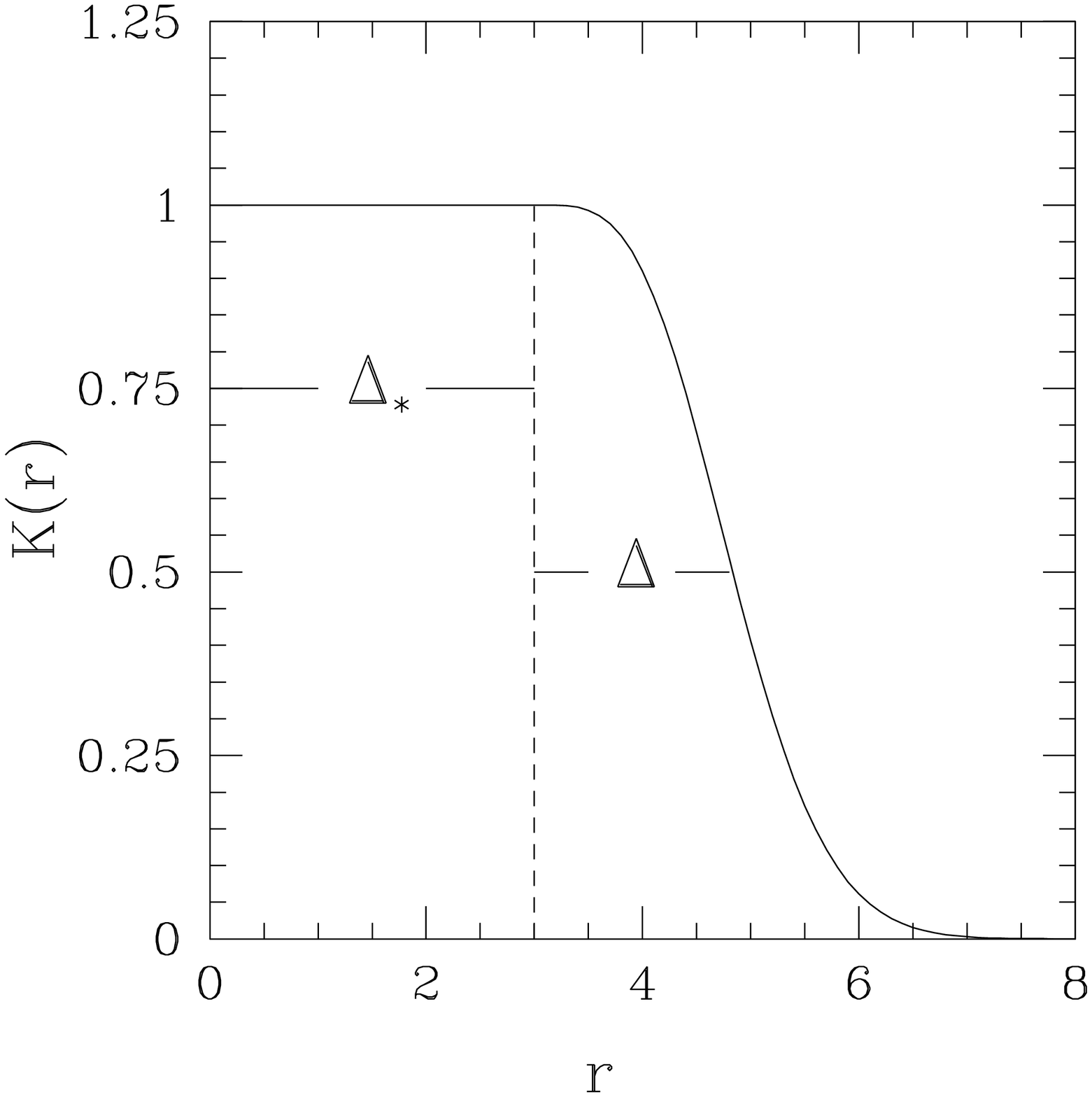,width=7.5cm}}
\caption{\label{new_param}\small This figure shows the
behaviour of $K(r)$ given by (\ref{new_curv_param}),
indicating explicitly the meaning of the two parameters
$\Delta$ and $\Delta_*$.} 
\end{figure}

\begin{figure}[t!]
\centerline{\psfig{file=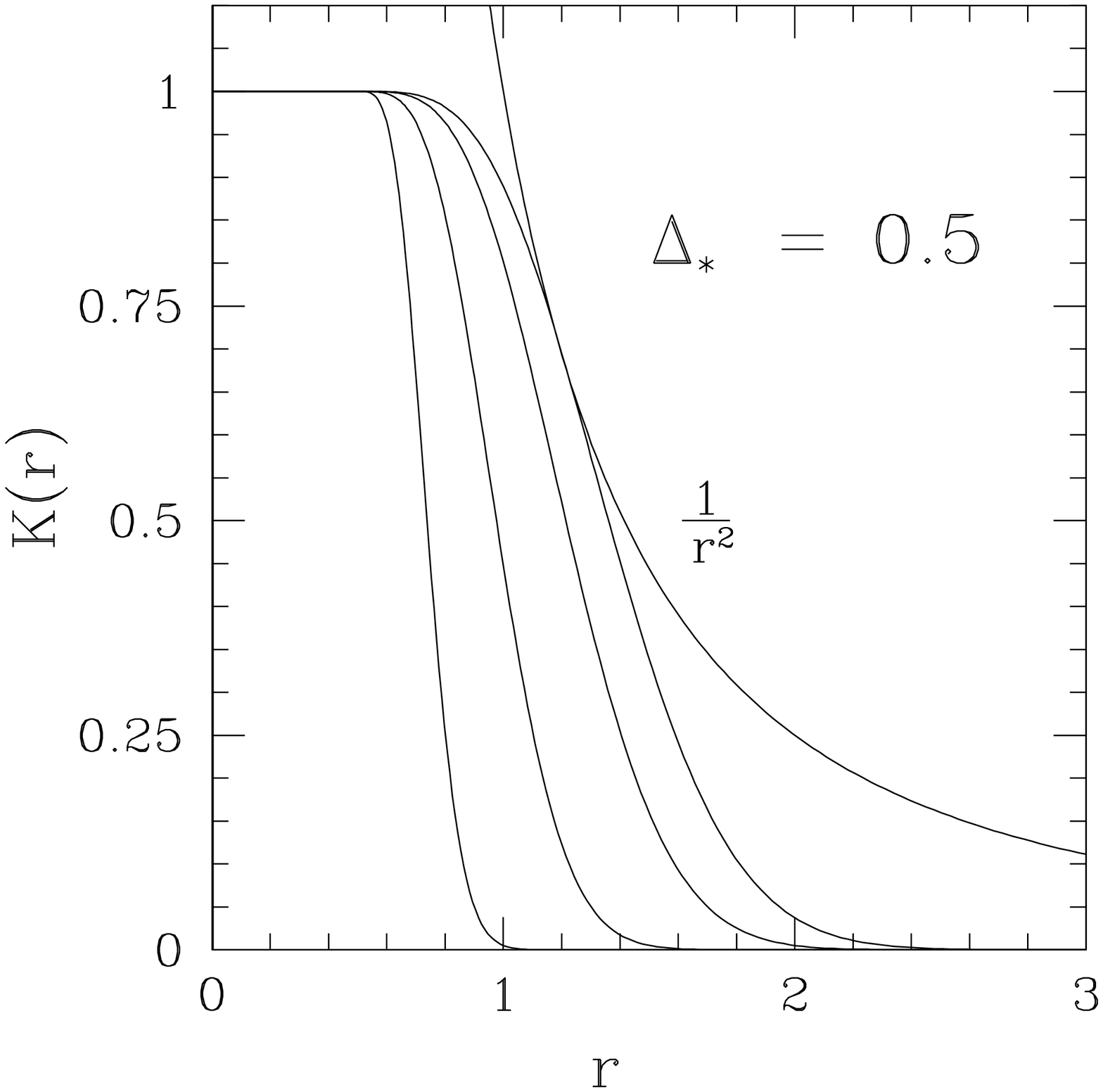,width=6.5cm}
            \psfig{file=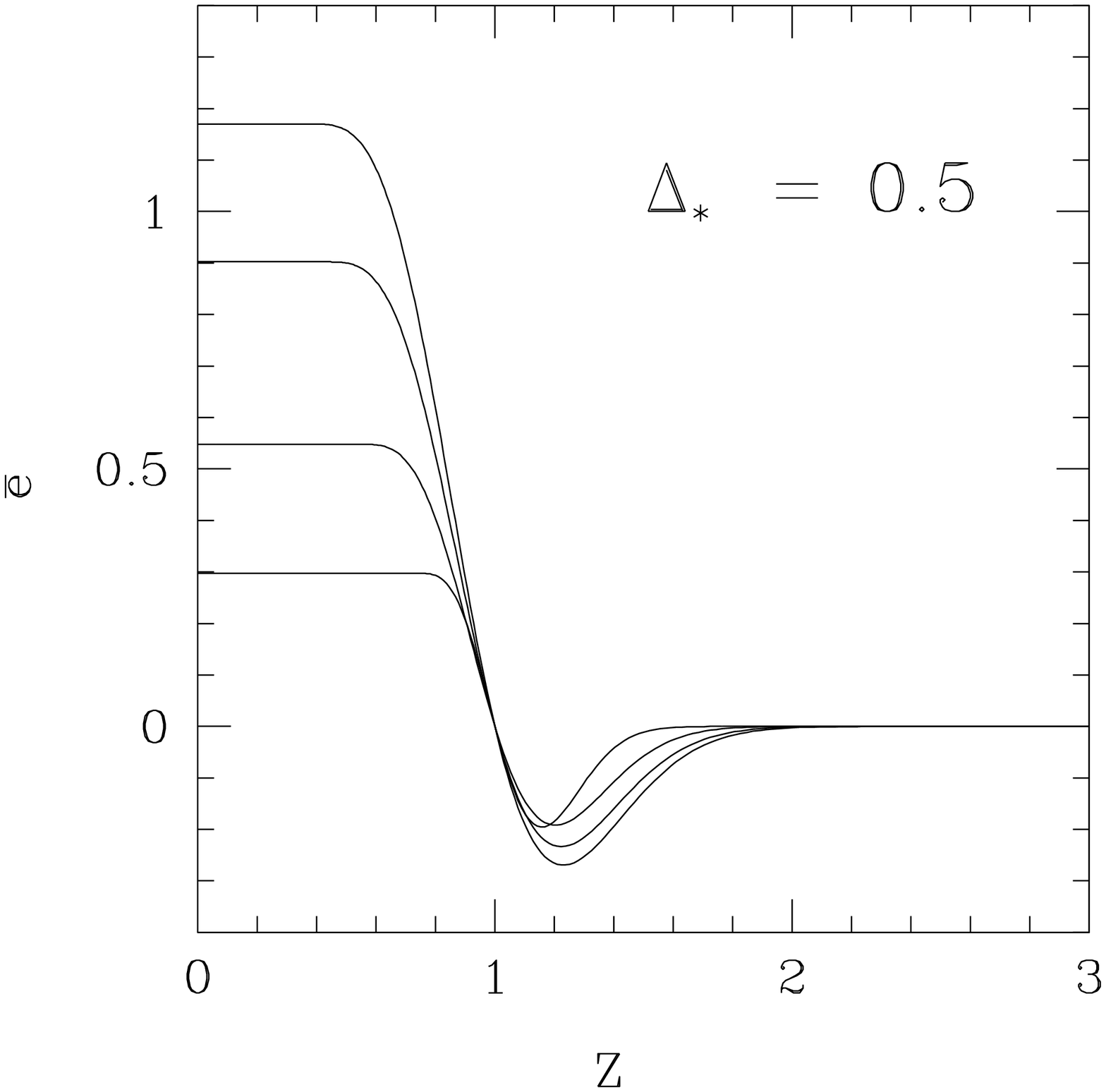,width=6.5cm}}
\centerline{\psfig{file=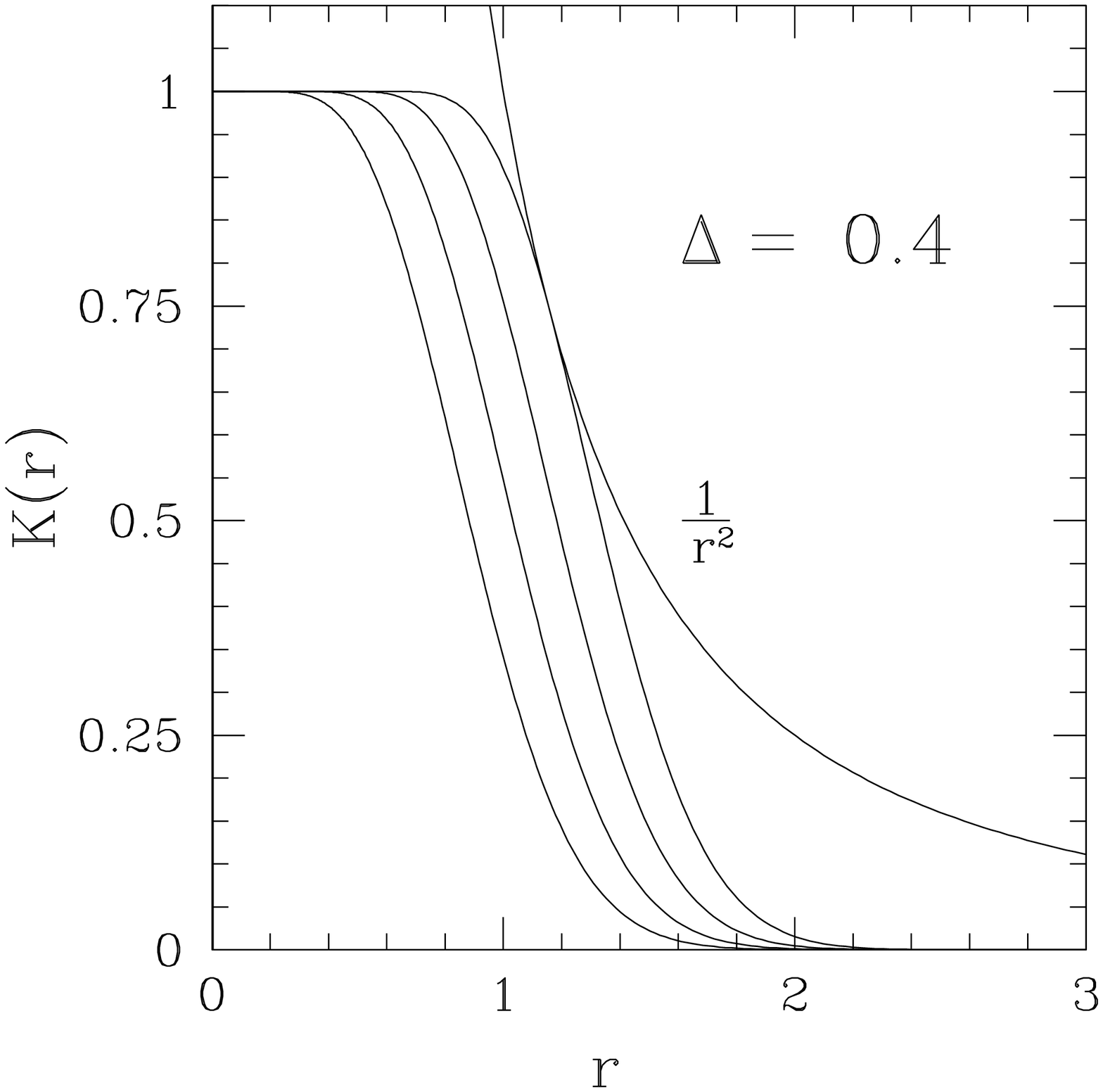,width=6.5cm}
            \psfig{file=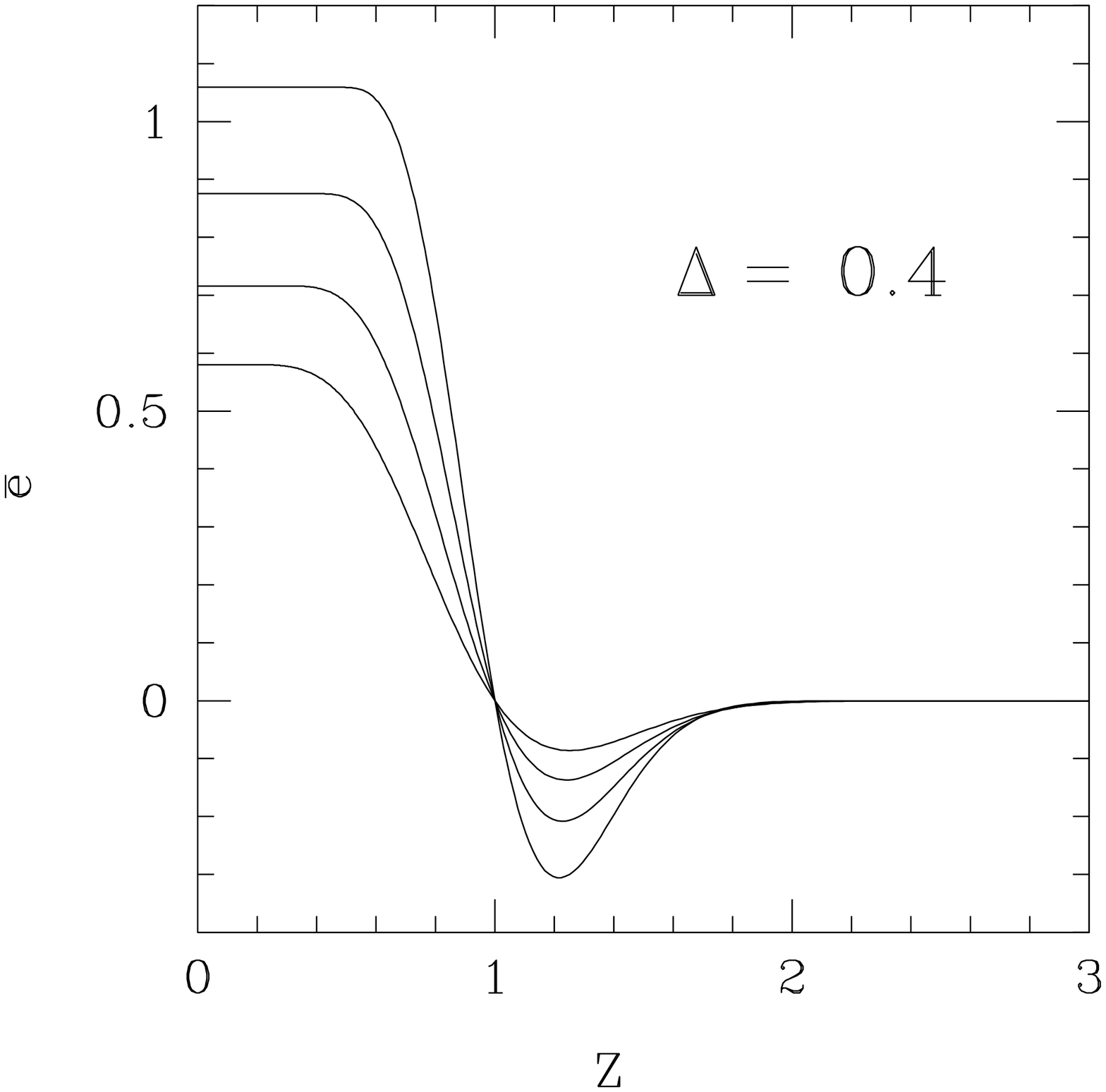,width=6.5cm}}
\caption{\label{profiles5}\small The left hand plots show
the curvature profile $K(r)$ for $\Delta_*$ constant and
varying $\Delta$ at the top row and for $\Delta$ constant
and varying $\Delta_*$ at the bottom. The corresponding
behaviour of the energy density perturbation $\tilde{e}$
is shown in the right hand plots.} 
\end{figure}

Next we consider another parametrisation of curvature
that allows us to work with curvature profiles with a
sharper transition from $K=1$ to  $K=0$. The new
parametrisation is given by the following expression: 
\bea 
K(r) \ug \disp{\left\{
\begin{array}{lc}
\ \,\, 1 \hspace{5.65cm} \mr{if} \quad r\,\leq\,\Delta_*\\ \\
\disp{\lp 1 + \frac{(r-\Delta_*)^2}{2\Delta^2} \rp\exp\lp -
\frac{(r-\Delta_*)^2}{2\Delta^2}\rp} \quad \mr{if}
\quad r\,>\,\Delta_* \\
\end{array}
\right.} \label{new_curv_param} 
\eea 
and its first derivative is
\bea 
K^\prime(r) \ug \disp{\left\{
\begin{array}{lc}
\ \,\, 0 \hspace{5.65cm} \mr{if} \quad r\,\leq\,\Delta_*\\ \\
\disp{-\,\frac{(r-\Delta_*)^3}{2\Delta^4} \exp\lp -
\frac{(r-\Delta_*)^2}{2\Delta^2}\rp} \ \ \qquad \mr{if}
\quad r\,>\,\Delta_* \\
\end{array}
\right.} 
\eea 
In figure \ref{new_param} we present the parametrisation
given by (\ref{new_curv_param}), indicating explicitly
the two independent parameters $\Delta$ and $\Delta_*$:
$\Delta_*$ specifies the radius of the plateau with
$K(r)=1$, $\Delta$ describes the sharpness of the
transition region. In particular, when $\Delta_*=0$ the
curvature profile coincides with the previous case for
$\alpha\ug1$. Applying condition (\ref{cond3}) we obtain
a quadratic equation to calculate the value of $r_0$, that
we solve numerically using the Newton-Raphson method \cite{Num_Rec}. The maximum value of $\Delta_*$ satisfying
condition (\ref{cond1}) is $1$, while the minimum value of
$\delta$, obtained when $\Delta=0$, is $\delta_{\rm{min}} \ug
2\Delta_*^2/3N^2$.

In Figure \ref{profiles5}, the top left plot shows the
dependence of $K(r)$ on $\Delta$ with $\Delta_*$ kept
fixed, while the bottom left plot shows the dependence of
$K(r)$ on $\Delta_*$ with $\Delta$ kept fixed. In the
right hand plots, we show the corresponding profiles of
$\tilde{e}(r)$. 
 
\section{Numerical tests and calculations}

\subsection{Numerical tests}
\begin{figure}[t!]
\centerline{\psfig{file=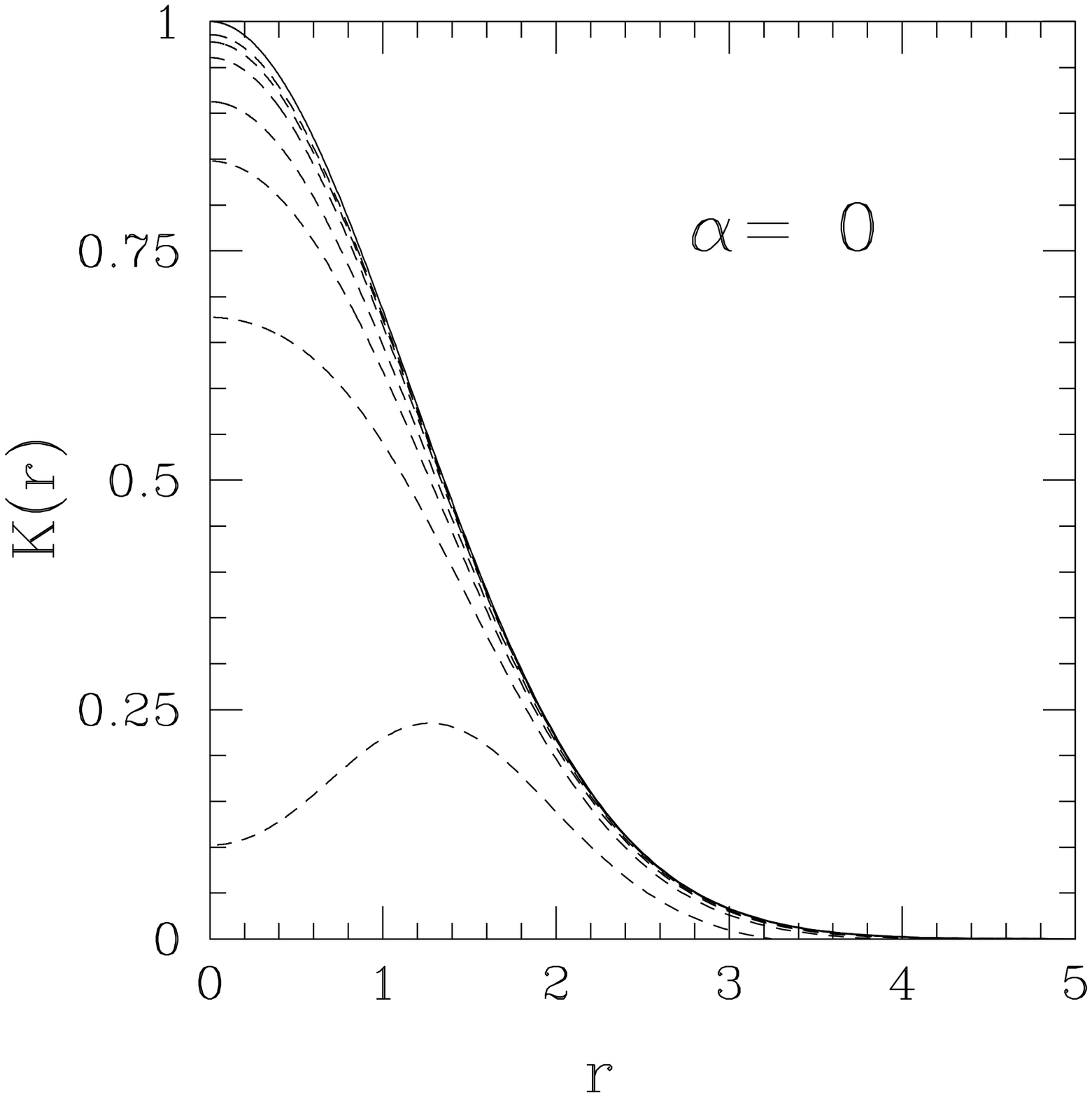,width=6.5cm}
            \psfig{file=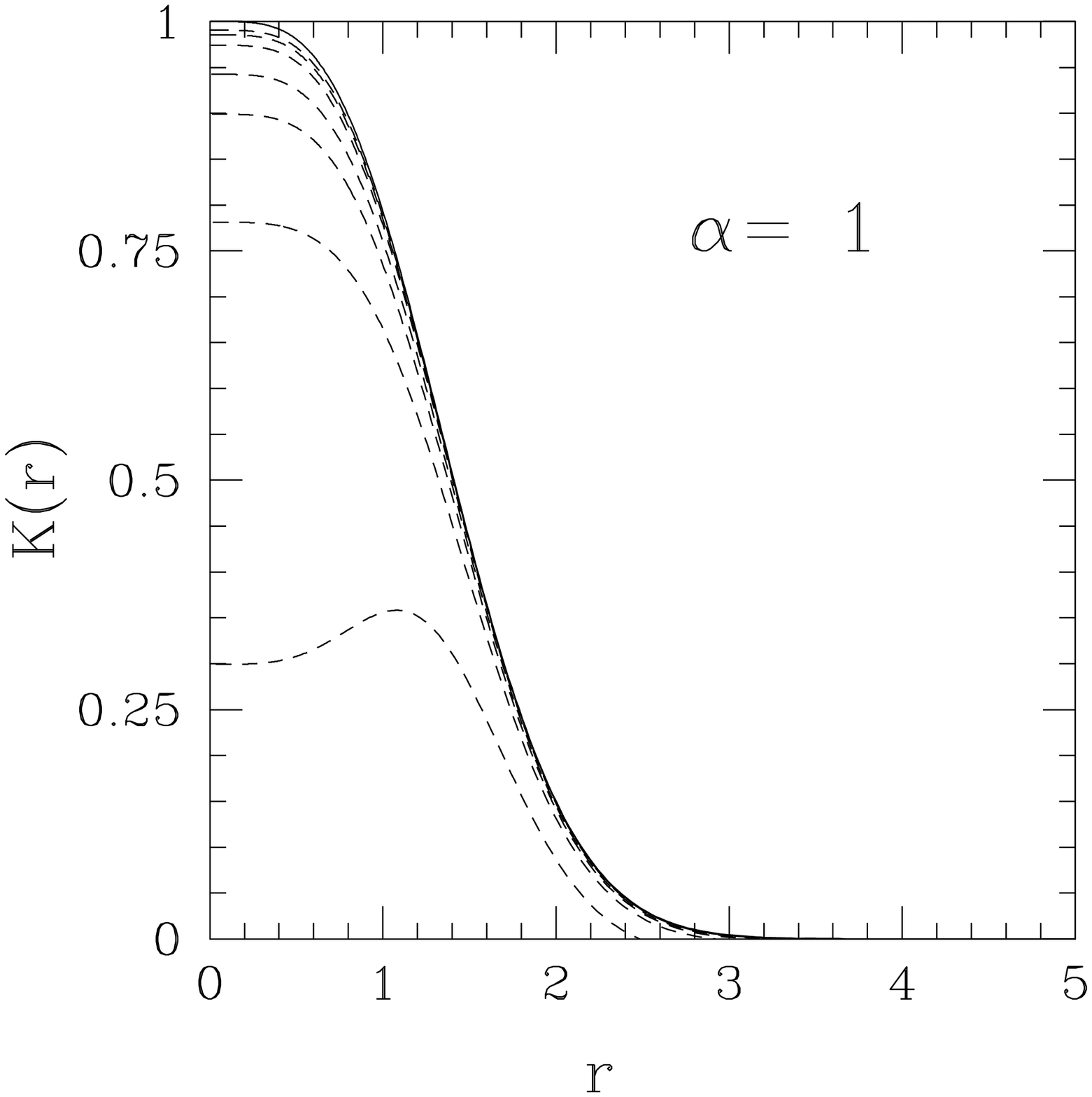,width=6.5cm}}
\caption{\label{test1}\small These plots show the
comparison between the analytical (solid line) and the
numerical (dashed line) profiles of curvature given by
(\ref{curv_profile}). The different dashed lines
correspond to $N=1,2,3,4,6,8,10$ with the higher curves
corresponding to the higher values of $N$. The profiles
in the left hand plot are characterised by ($\alpha\ug0$,
$\Delta\ug1.15$), and the profiles in the right hand plot
by ($\alpha\ug1$, $\Delta\ug0.77$).}
\end{figure}
Next we discuss the numerical tests performed so as to be
confident that the perturbation profile $K(r)$ has been
introduced consistently into the code. For this we
analysed the accuracy of our first order approach (first
order with respect to $\epsilon \ug /N^2$). In other
words, imposing a perturbation with a length scale much
larger then the cosmological horizon $r_H$, we checked
whether $\epsilon$  is small enough to make higher order
terms negligible. A first requirement for the necessary
value of $N$  is that the comoving lengthscale of the
horizon $r_H$ should be sufficiently smaller than the
sharpness $\Delta$ of the curvature profile. This means
that $r_H\,<<\,\Delta$. From equation (\ref{r_def}), $r_H
\ug r_0/N$ which then gives the  condition, 
\beq 
N \,>>\, \frac{r_0}{\Delta}\,. 
\label{cond_N1}
\eeq

Using the curvature profile given by (\ref{curv_profile})
we have $r_0 \simeq \Delta$ for $0\leq\alpha\leq1$, while
using the curvature profile given by
(\ref{new_curv_param}) we have that $r_0 \simeq \Delta_*
+ \Delta$. Substituting these two relations into
(\ref{cond_N1}), we see that both conditions are
satisfied if 
\beq 
N\,>>\, 1+\frac{\Delta_*}{\Delta}\,. 
\label{cond_N2} 
\eeq
This expression is valid also for the curvature profile
given by (\ref{curv_profile}) if we put
$\Delta_*\ug0$. From this expression we can see that a
sharper profile, with $\Delta_*/\Delta\,\gg\,1$, requires
larger values of $N$. 

\begin{figure}[t!]
\centerline{\psfig{file=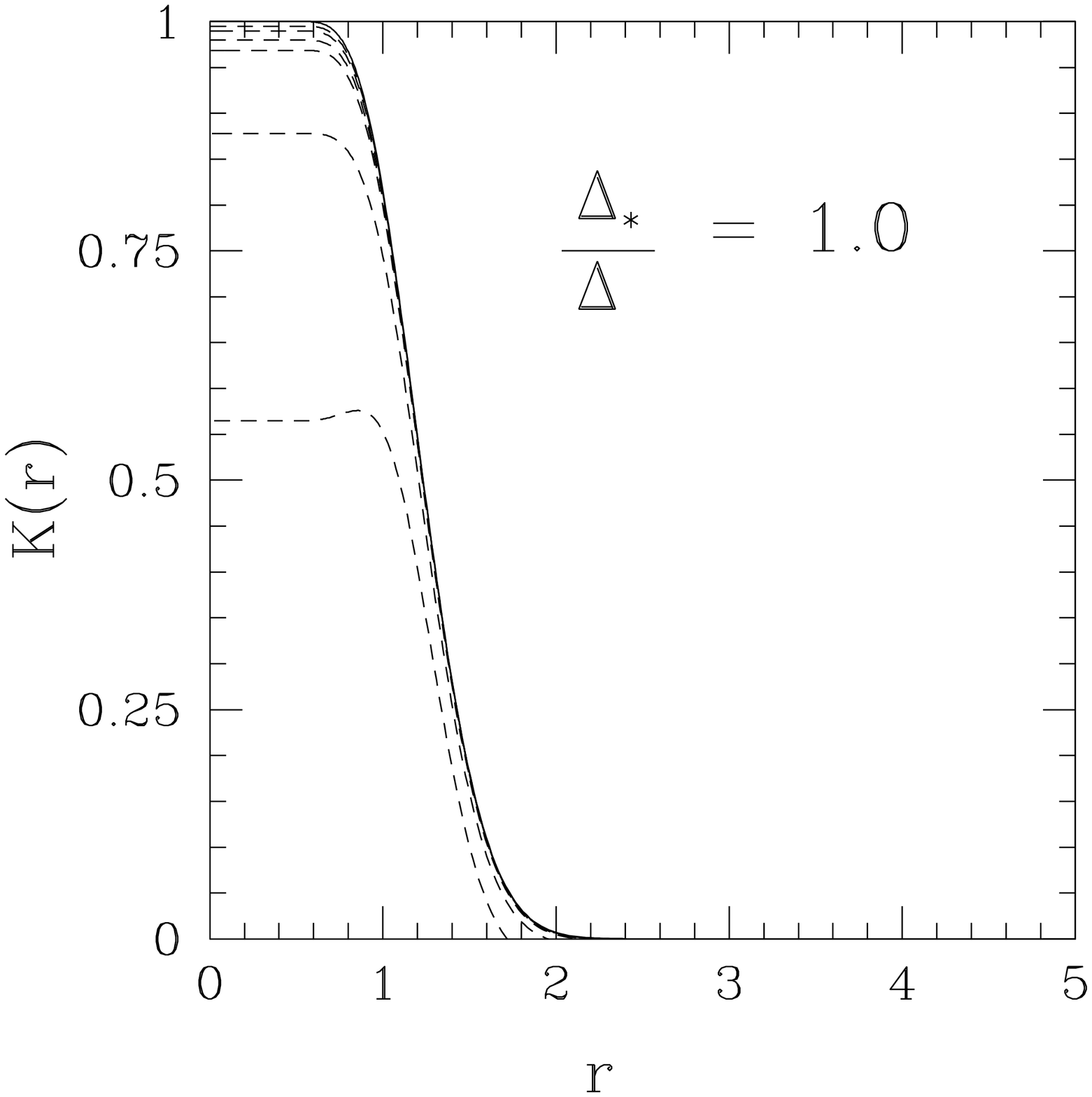,width=6.5cm}
        \psfig{file=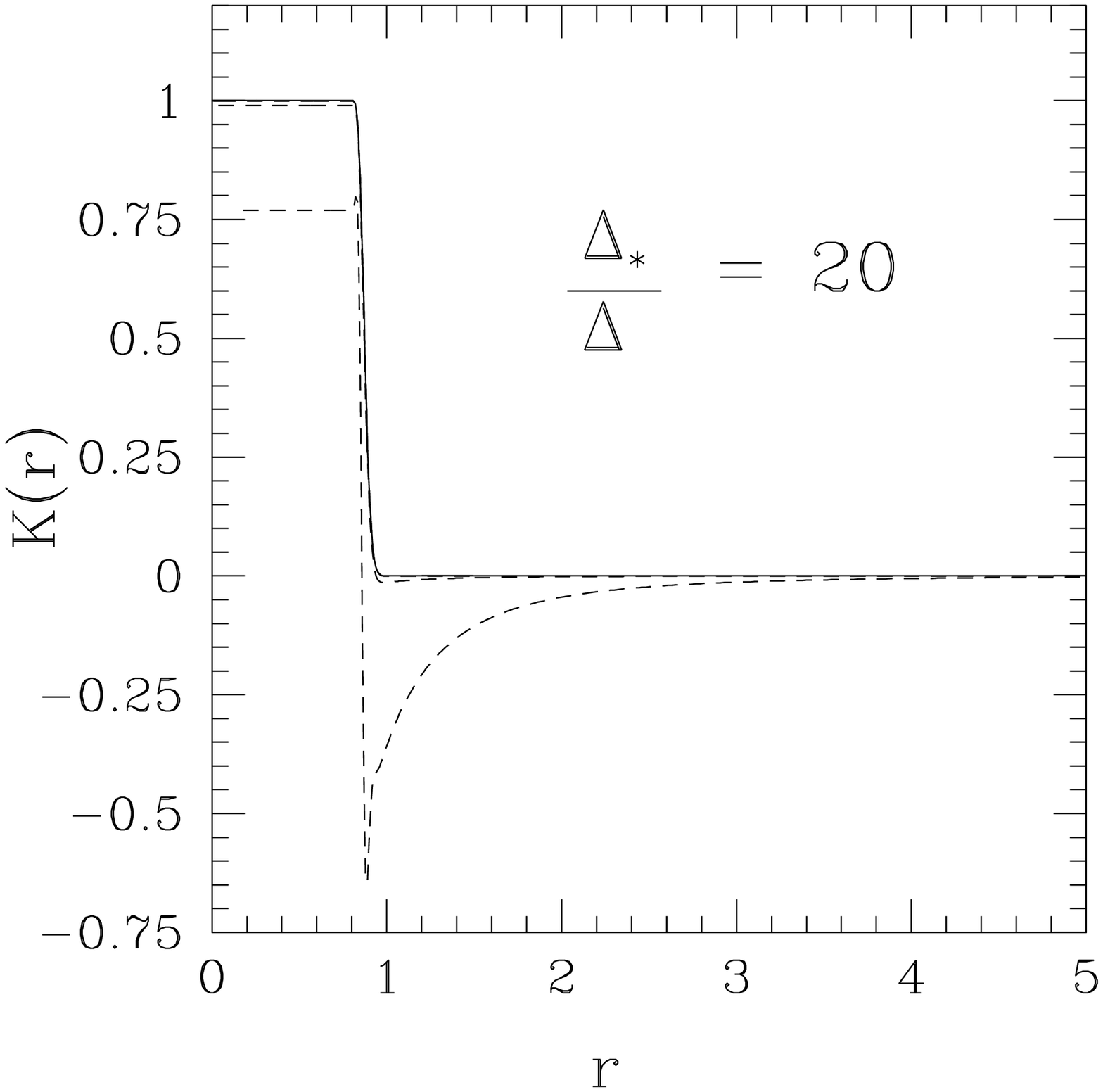,width=6.5cm}}
\caption{\label{test2}\small These plots show the comparison
between the analytical (solid line) and the numerical (dashed
lines) profiles of curvature given by (\ref{new_curv_param}). The
different dashed lines correspond to $N=1,2,4,5,7,10$ in the left
plot and $N=1,5,25$ in the right one, with the higher curves
corresponding to the higher values of $N$. The profiles in the
left hand plot are characterised by $\Delta_*/\Delta=1$, and the
profiles in the right hand plot by $\Delta_*/\Delta=20$.}
\end{figure}

For each set of initial input parameters we apply several
numerical tests. Here we present four sample cases that
demonstrate how these tests work. The main test is based
on equation (\ref{K_Gamma2}), which is used to rederive
$K(r)$ after calculating the initial conditions for the
various quantities. Expression (\ref{K_Gamma2}) is valid
when the higher order terms are negligible and so we
expect to find some difference between the profiles of
$K(r)$ calculated with (\ref{K_Gamma2}) and the original
ones introduced analytically to describe the initial
conditions. In Figure \ref{test1} we plot the result of
this test using $K(r)$ given by (\ref{curv_profile}),
with $\alpha\ug0$ on the left and $\alpha\ug1$ on the
right. In Figure \ref{test2} we plot $K(r)$ given by
(\ref{new_curv_param}), using the ratio $\Delta_*/\Delta$
equal to $1$ in the left panel, and equal to $20$ in the
right panel. In all of these plots the solid line shows
the analytical profile of $K(r)$, while the dashed lines
corresponds to the profiles of $K(r)$ calculated with
(\ref{K_Gamma2}) using different values of $N$.  The
first value used was $N\ug1$, that we know is not large
enough, and this is also clear from the plots; we then
increased $N$. In the first three cases, corresponding to
curvature profiles with $\Delta_*/\Delta\,\leq\,1$, we
found that $N\ug10$ is large enough to achieve acceptable
precision. In the last case, when $\Delta_*/\Delta\ug20$,
we found that $N=25$ is sufficiently large.  Because we
did not see any significant difference between the
analytical and numerical values of $K(r)$, for the
highest values of $N$ plotted in Figure \ref{test1} and
\ref{test2} we assume that these values of $N$ are
satisfactory. To be more certain of this we also made
another test, analysing  numerically the  evolution of
configurations corresponding to the curvature profiles
$K(r)$ presented  in Figure \ref{test1} and
\ref{test2}. We have verified that the values of $N$
indicated by the previous test do give essentially the
same numerical evolution as for larger values of $N$ (a
standard test for numerical computations). 

To complete our tests we used the constraint equation
(\ref{Gamma}) to estimate the precision of the code and
we found that, with the number of gridpoints that we used
in our simulations, the numerical scheme has accuracy
better than one part in $10^4$ (by the numerical scheme
here we mean the code itself plus the initial
conditions.) 

\subsection{Description of the calculations}
The calculations of black hole formation performed with
initial conditions specified with a curvature profile,
show similar hydrodynamical features as those presented
in \cite{Musco}, where the same numerical technique was
used. We present here four sample cases: two correspond
to black hole formation, where the observer time slicing
has been used, and two correspond to no black hole
formation, where we have used the cosmic time
slicing. Each case is characterised by a particular value
of the density perturbation amplitude $\delta$ and it is
useful to introduce at this stage a measure of $\delta$
which is independent of the initial scale used when
imposing the initial conditions. Looking at expression
(\ref{delta_s}), we can see that in the case of constant
$\gamma$, if we define
$\tilde{\delta}\,\equiv\,N^2\delta$, the expression
becomes time independent, giving in the radiation epoch 
\beq 
\tilde{\delta} \ug \frac{2}{3}K(r_0)r_0^2\,. 
\label{delta_tilda} 
\eeq 
Comparing this value with the value of $\delta$ at
horizon crossing, we find that it is of the same order
(with a small correction due to the non linear growth of
the energy density near to horizon crossing). Therefore
for the rest of the discussion we use $\tilde{\delta}$ as
the amplitude of density perturbations. The dividing line
between perturbations that collapse to form black holes
and perturbations that disperse into the background is
characterised by a threshold amplitude
$\tilde{\delta}_c$. This value was estimated numerically
by making a converging sequence of numerical calculations
until a satisfactory precision in determination of
$\tilde{\delta}_c$ was achieved.  

\begin{figure}[t!]
\centerline{\psfig{file=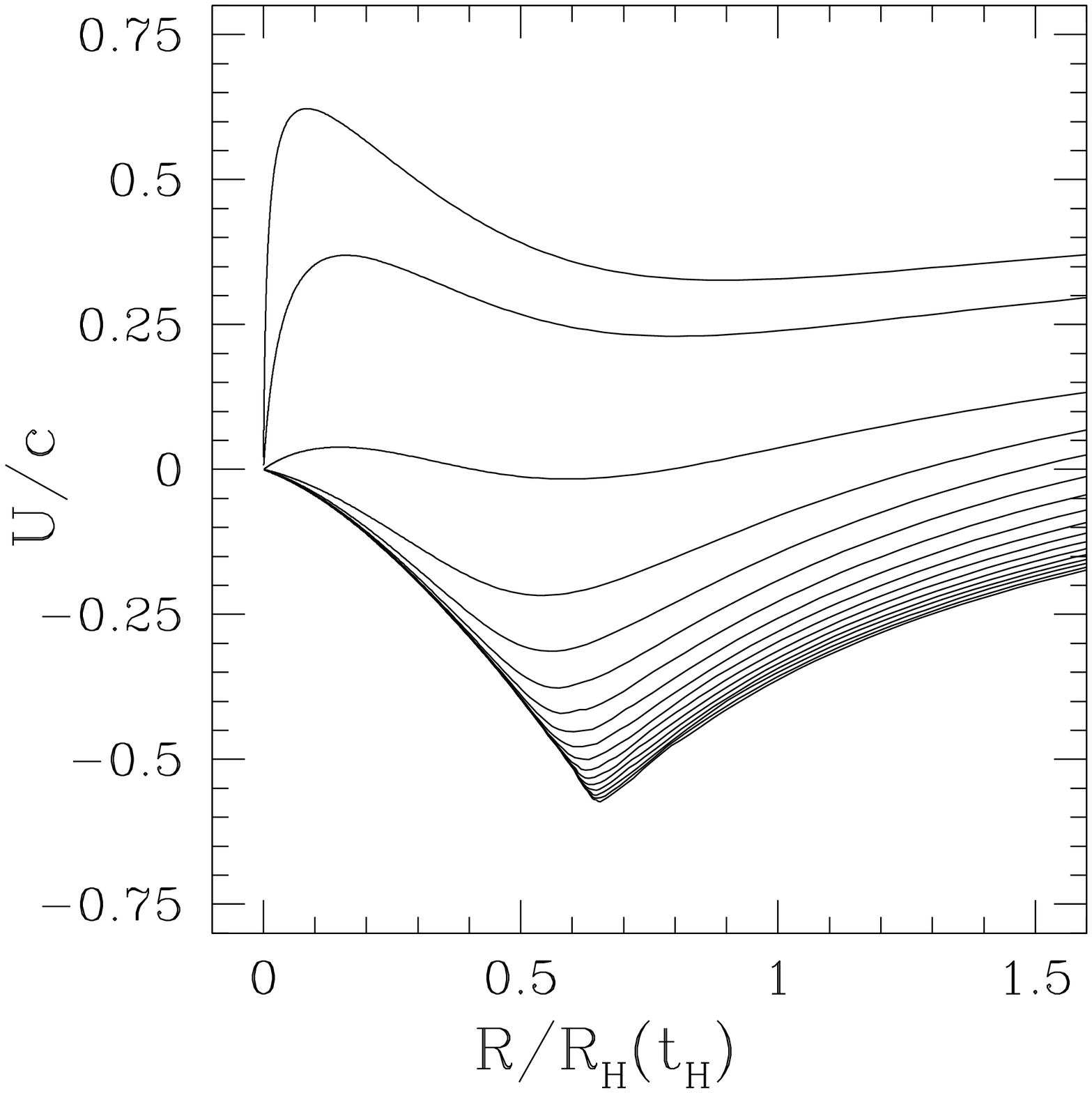,width=6.5cm}
            \psfig{file=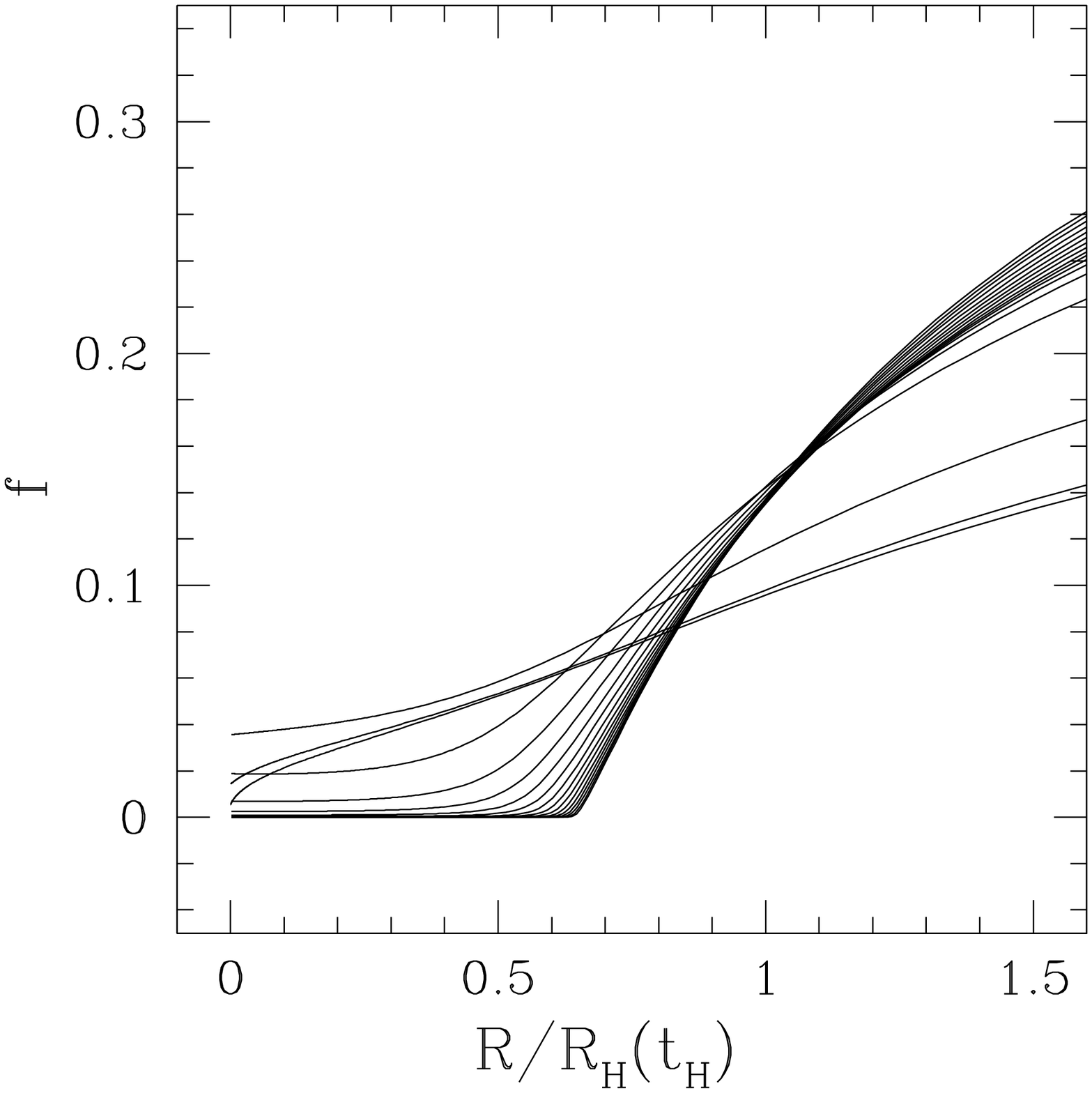,width=6.5cm}}
\centerline{\psfig{file=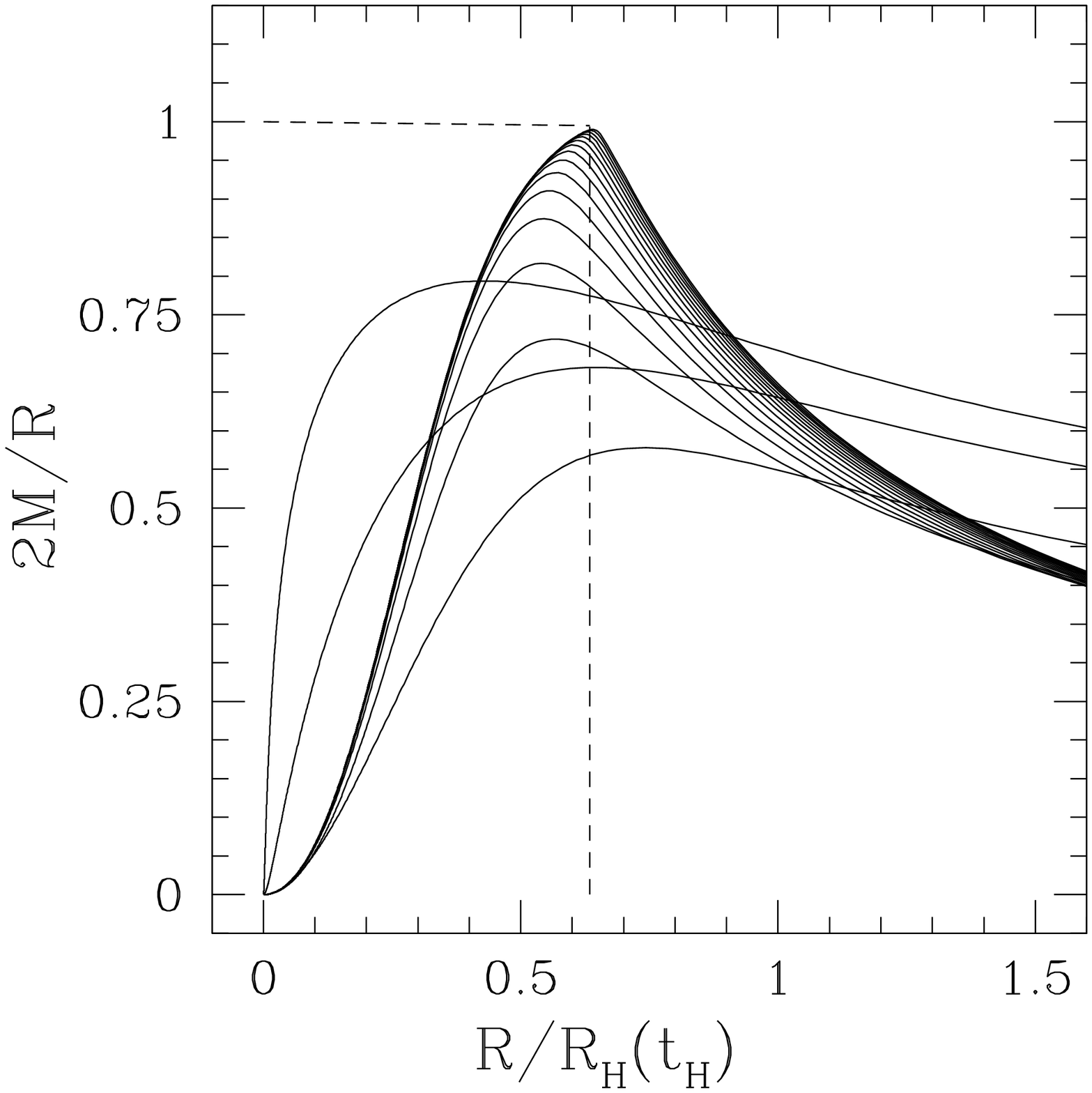,width=6.5cm}
            \psfig{file=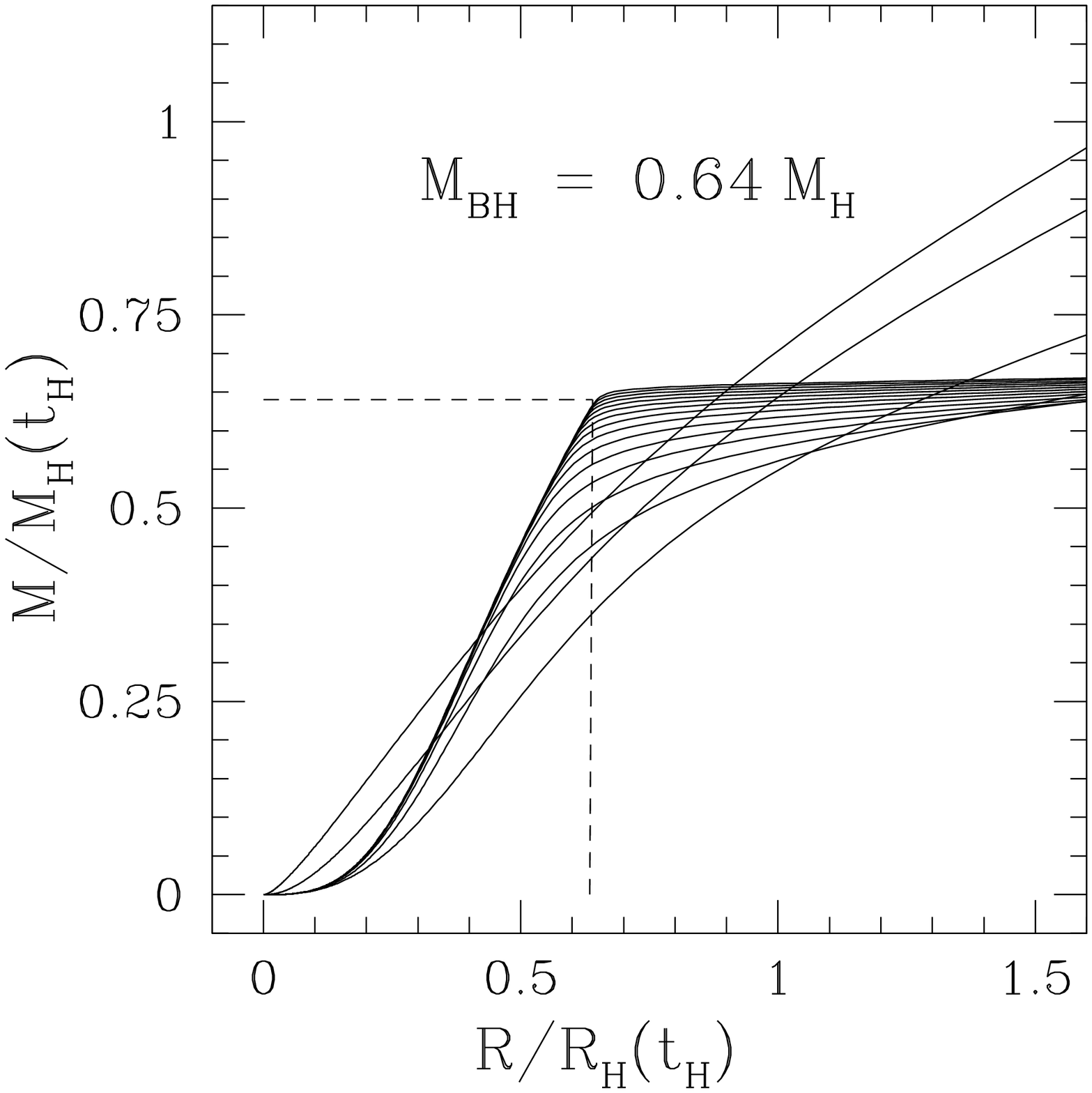,width=6.5cm}}
\caption{\label{bh_form1}\small A typical evolution leading to
black hole formation: the initial curvature profile used is
characterised by $\alpha=0$ and $\Delta=1.02$, which give
$(\tilde{\delta}-\tilde{\delta}_c)\,\ug1.3\,\times\,10^{-2}$. The
upper plots show the profile of radial velocity $U/c$ and lapse
$f$ at different times, while the bottom panels show the
corresponding profiles of $2M/R$ and mass $M$.}
\end{figure}

\begin{figure}[t!]
\centerline{\psfig{file=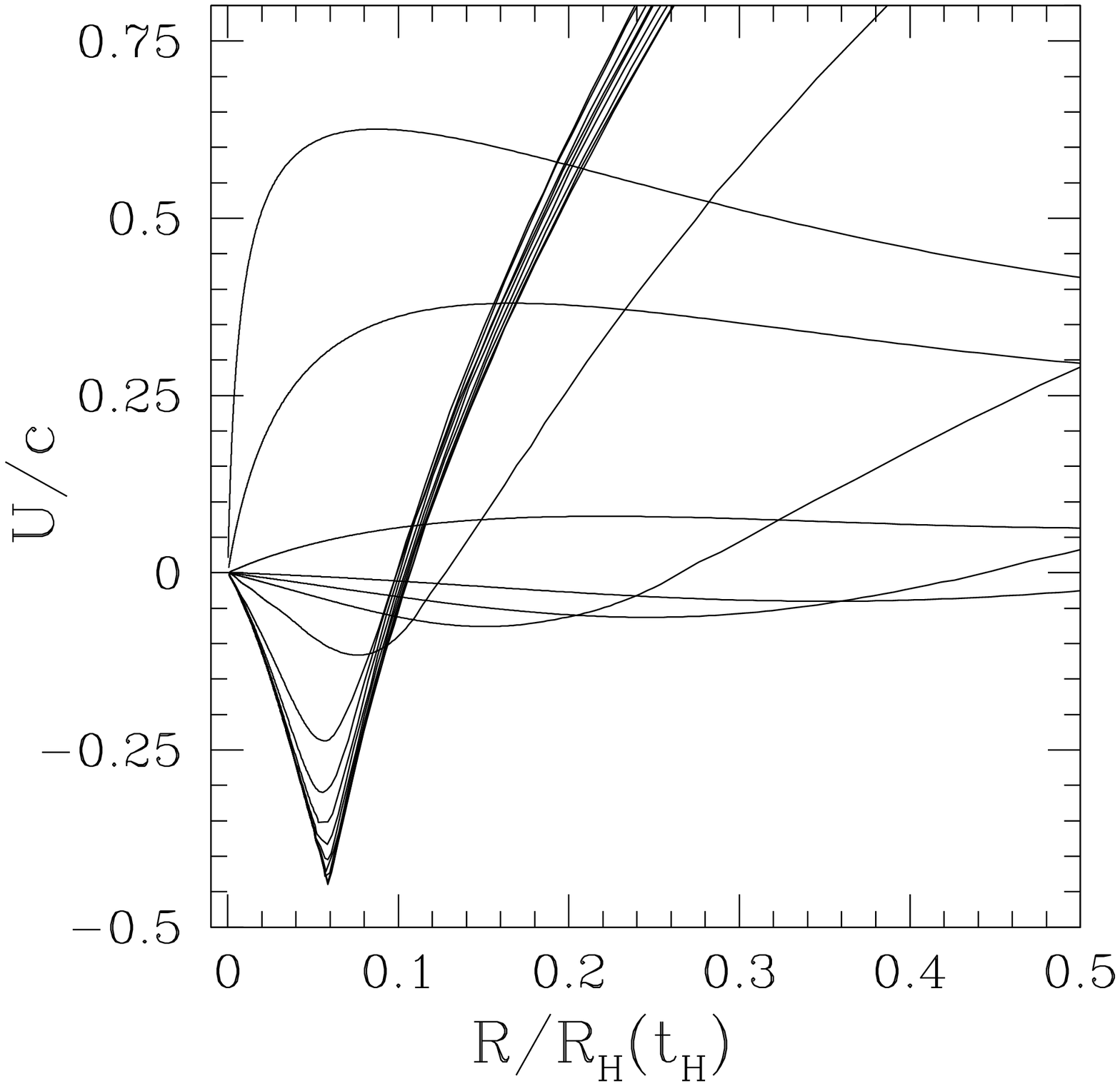,width=6.5cm}
            \psfig{file=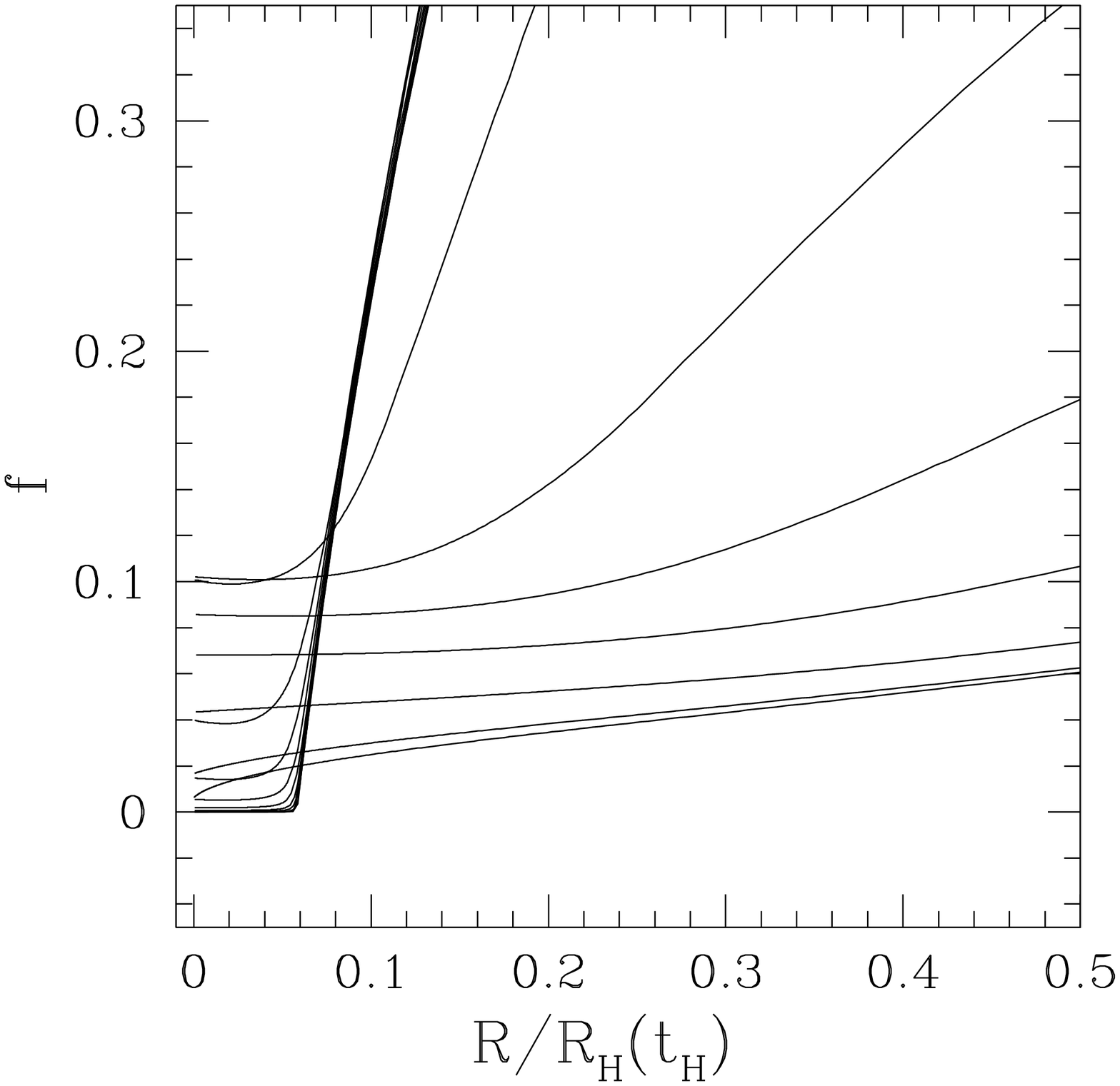,width=6.5cm}}
\centerline{\psfig{file=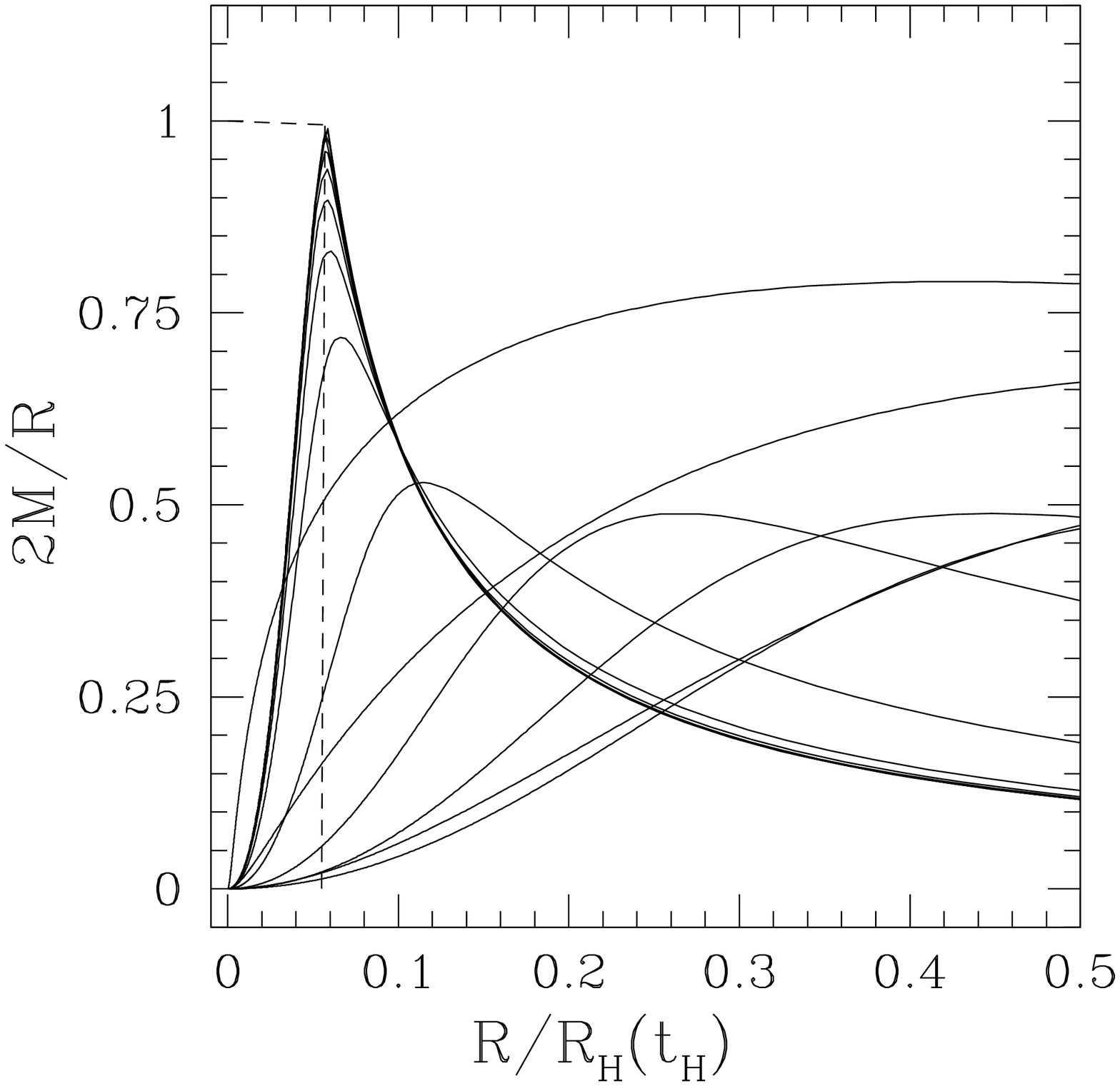,width=6.5cm}
            \psfig{file=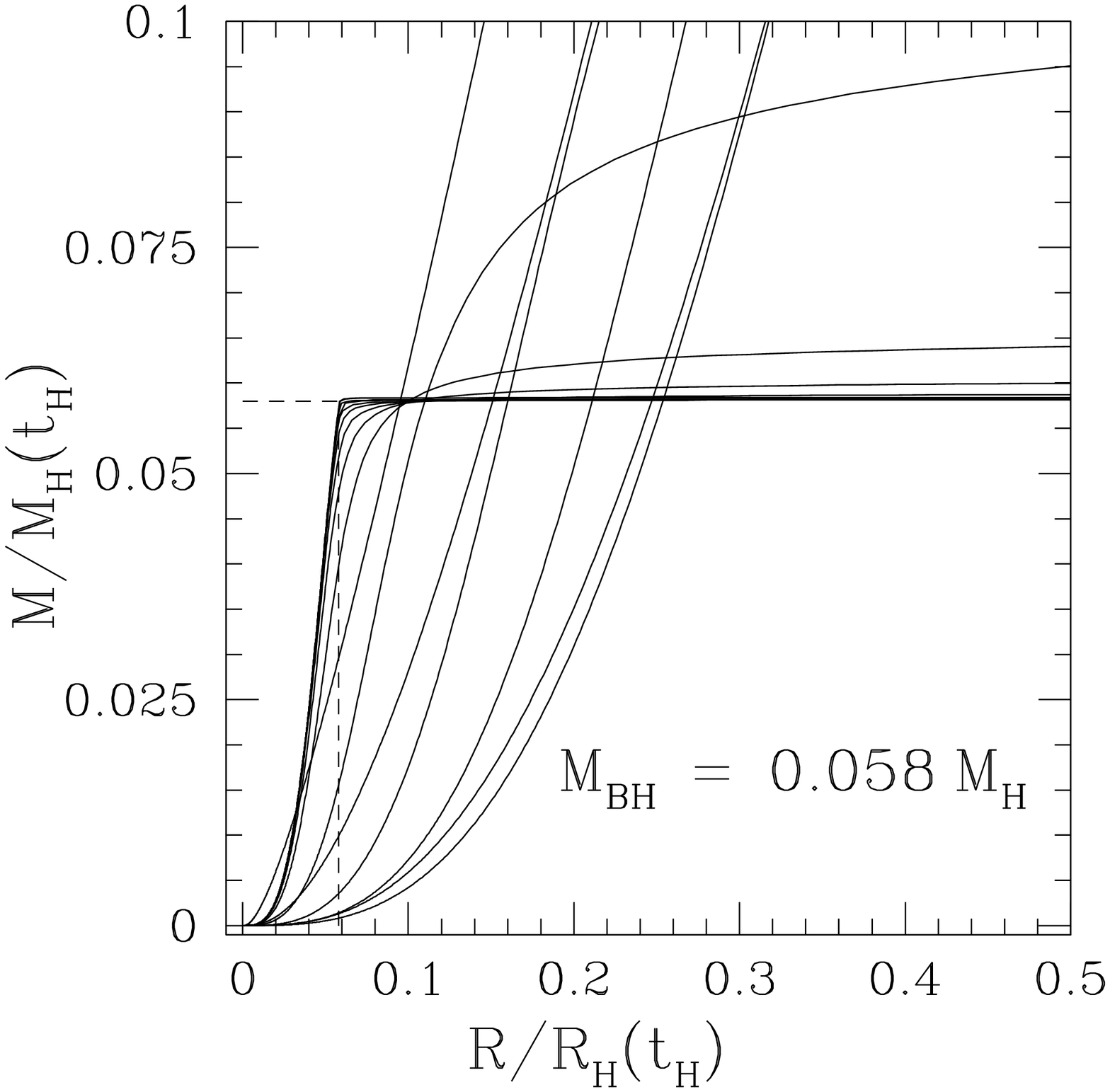,width=6.5cm}}
\caption{\label{bh_form2}\small A typical evolution leading to
black hole formation: the initial curvature profile used is
characterised by $\Delta_*=0.8$ and $\Delta=0.18$, which give
$(\tilde{\delta}-\tilde{\delta}_c)\,\ug \,1.8\times\,10^{-3}$. The
upper plots show the profile of radial velocity $U/c$ and lapse
$f$ at different times, while the bottom panels show the
corresponding profiles of $2M/R$ and mass $M$.}
\end{figure}

The observer time coordinate $u$ is defined by 
\beq 
f\,du \ug a\,dt \,-\, b\,dr 
\eeq 
where $f$ is the lapse in the new coordinate system. The
metric is then   
\beq 
ds^2 \ug - f^2\,du^2 \,-\, 2fb\,dr\,du \,+\,
R^2\,(d\theta^2 \,+\,sin^2\theta\,d\phi^2)\,.  
\eeq 
When using observer-time slicing, black hole formation
occurs when both the lapse $f$ and $\Gamma+U$ tend to
zero (see \cite{Musco}). These conditions are reached
asymptotically and correspond to the formation of the
event horizon, seen by a distant observer with an
infinite redshift. 

We illustrate the two examples of black hole formation,
shown in Figures \ref{bh_form1} and \ref{bh_form2}, by
plotting profiles of the radial velocity $U$, lapse $f$,
$2M/R$ ratio and mass $M$ as functions of the
circumferential radius $R$ at different times. These show
the formation of black holes for different values of
($\tilde{\delta} - \tilde{\delta}_c$) with the smaller
value giving rise to the lower mass black hole as
expected \cite{Nadezhin,Niemeyer1,Hawke,Musco}. 

The upper left hand plots of these two figures show the
behaviour of the radial velocity $U$. Here the time
sequence starts from the top curve and continues towards
the bottom one. Note that, after a certain time
appreciate how $U$ becomes negative in the central
region, corresponding to the region where the black hole
forms. In particular the event horizon, as can be seen
clearly looking at the plots for $2M/R$, forms at the
minimum of the radial velocity profile. Outside this
region there is first an intermediate shell with negative
velocity corresponding to matter accreting into the black
hole, then there is a region of positive velocity
corresponding to the rest of the expanding universe. 

The upper right hand plots show the behaviour of the
lapse $f$ with the time sequence of the curves given by
the decreasing central value of $f$.  One can sees that a
plateau forms in the central region, where the black hole
is forming, and that the lapse tends asymptotically to
zero there. The ``freezing'' of the evolution, as
mentioned above, is one of the fundamental  criteria for
black hole formation. The other important feature to
ensure us about black hole formation is the trapped
surface condition ($\Gamma+U \rightarrow 0$) which
corresponds to $2M/R \rightarrow 1$. In the left bottom
plots $2M/R$ is shown asymptotically tending to $1$ at
the maximum of the curve. This determines the location
where the event horizon will form. In this case the time
sequence of the curves is given at first by the
decreasing values of $2M/R$ at the right edge of the
plot, and then by the increasing of the maximum of these
curves towards $1$. 

Finally the bottom right hand plots show the
corresponding behaviour of the mass $M$, indicating the
value of the mass of the black hole.At late times, one
sees the convergence of the curves. 
In figure \ref{bh_form1} and \ref{bh_form2} the
circumferential radius $R$ and mass $M$ have been
normalized with respect to the values for the
cosmological horizon scale at the horizon crossing time. 

\begin{figure}[t!]
\centerline{\psfig{file=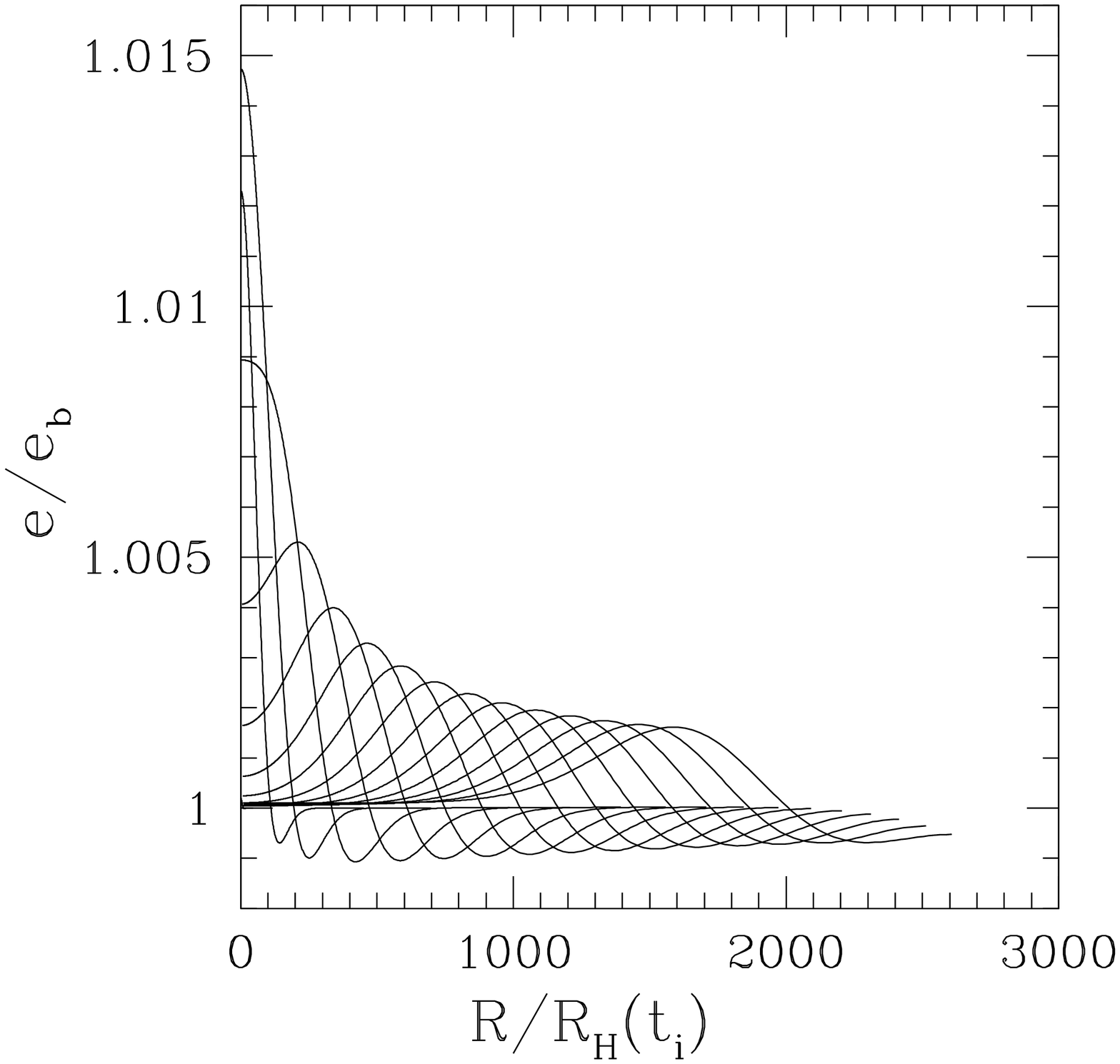,width=6.5cm}
            \psfig{file=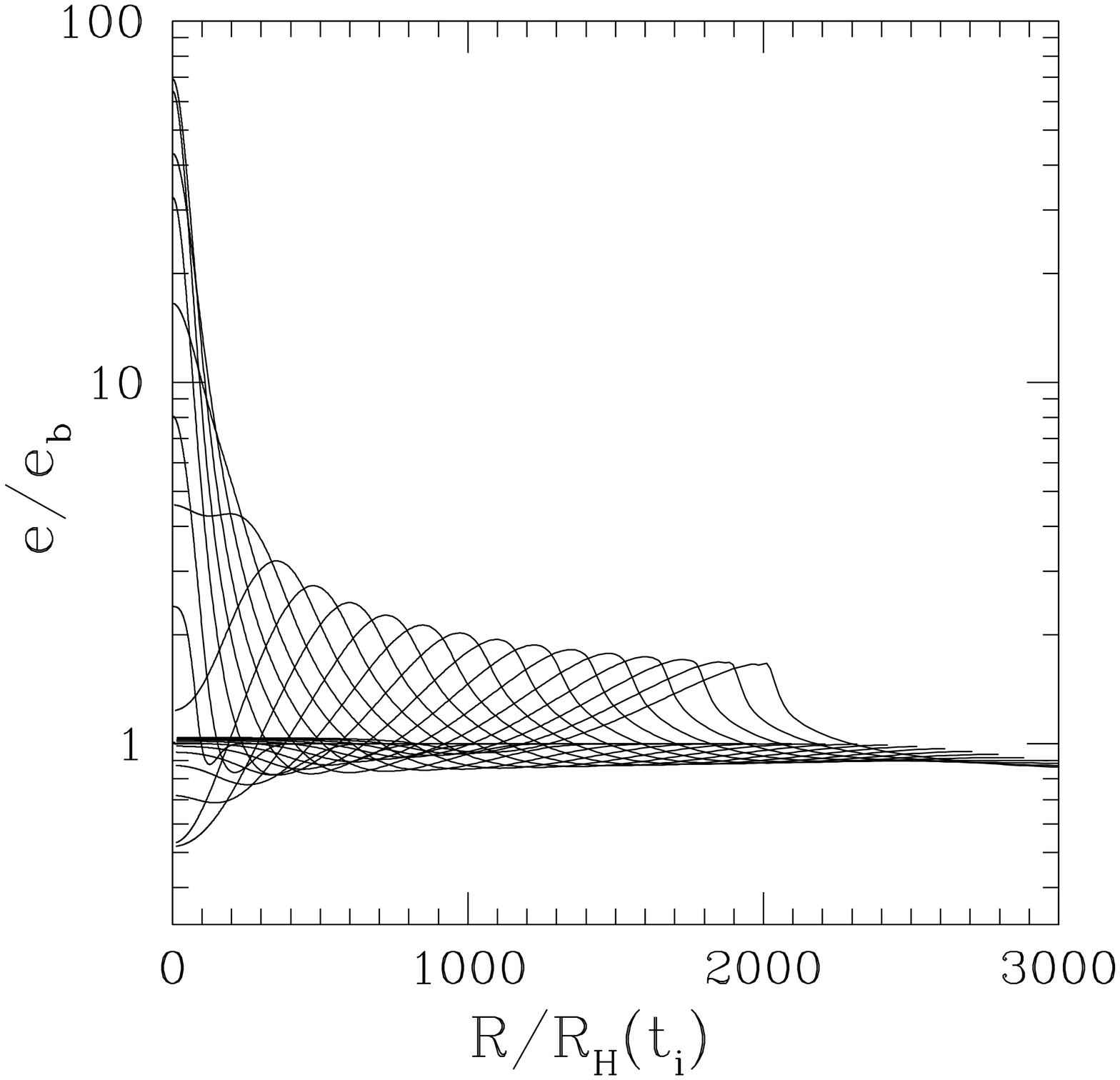,width=6.5cm}}
\caption{\label{Sub_en}\small The left hand plot shows a typical
evolution of the energy density profile of a perturbation with
$\tilde{\delta}\ll\tilde{\delta}_c$. In this case we have used
expression (\ref{curv_profile}) with $\alpha=0$, $\Delta=0.1$ and
$N=10$, corresponding to $\tilde{\delta}=4.6 \cdot
10^{-3}$. The right hand plot show the typical evolution of a
collapse when
$\tilde{\delta}\lesssim\tilde{\delta}_c$. In this case we have
used expression (\ref{new_curv_param}) with $\Delta_*=0.5$,
$\Delta=0.37$ and $N=10$, corresponding to
$\tilde{\delta}=0.49$.}
\end{figure}

Results from two representative calculation with
$\tilde{\delta}<\tilde{\delta}_c$ (using cosmic time) are
shown in Figure \ref{Sub_en}, where we have plotted the
energy density normalised with respect to the current
background value, as a function of $R/R_H(t_i)$. The left
hand plot corresponds to a small curvature perturbation
far from the threshold $\tilde{\delta}_c$ with
$\tilde{\delta} \ug 4.46 \cdot 10^{-3}$ and
$(\tilde{\delta}-\tilde{\delta}_c) \, \ug \, -0.45$. In
this case negative velocities are never reached:
initially the slower expansion of the central region
proceeds more slowly than the expansion of the
background. After horizon crossing, the perturbation is
damped and compression a wave propagates outwards through
the expanding medium. The other example shown in the
right hand plot shows a case with
$(\tilde{\delta}-\tilde{\delta}_c) \ug -2.67 \cdot
10^{-2}$. In this case the perturbation amplitude
$\tilde{\delta} \ug 0.49$ is close enough to
$\tilde{\delta}_c \ug 0.52$ to give rise to an initial
collapse of the central region. Because the gravitational
potential is not large enough to form a trapped surface,
as happens when $\tilde{\delta}>\tilde{\delta_c}$, the
collapsing matter bounces producing a compression wave
which expands outwards. These hydrodynamical features
were first seen in \cite{Nadezhin} and were then analysed
in detail in $\cite{Musco}$.

\begin{figure}[t!]
\centerline{\psfig{file=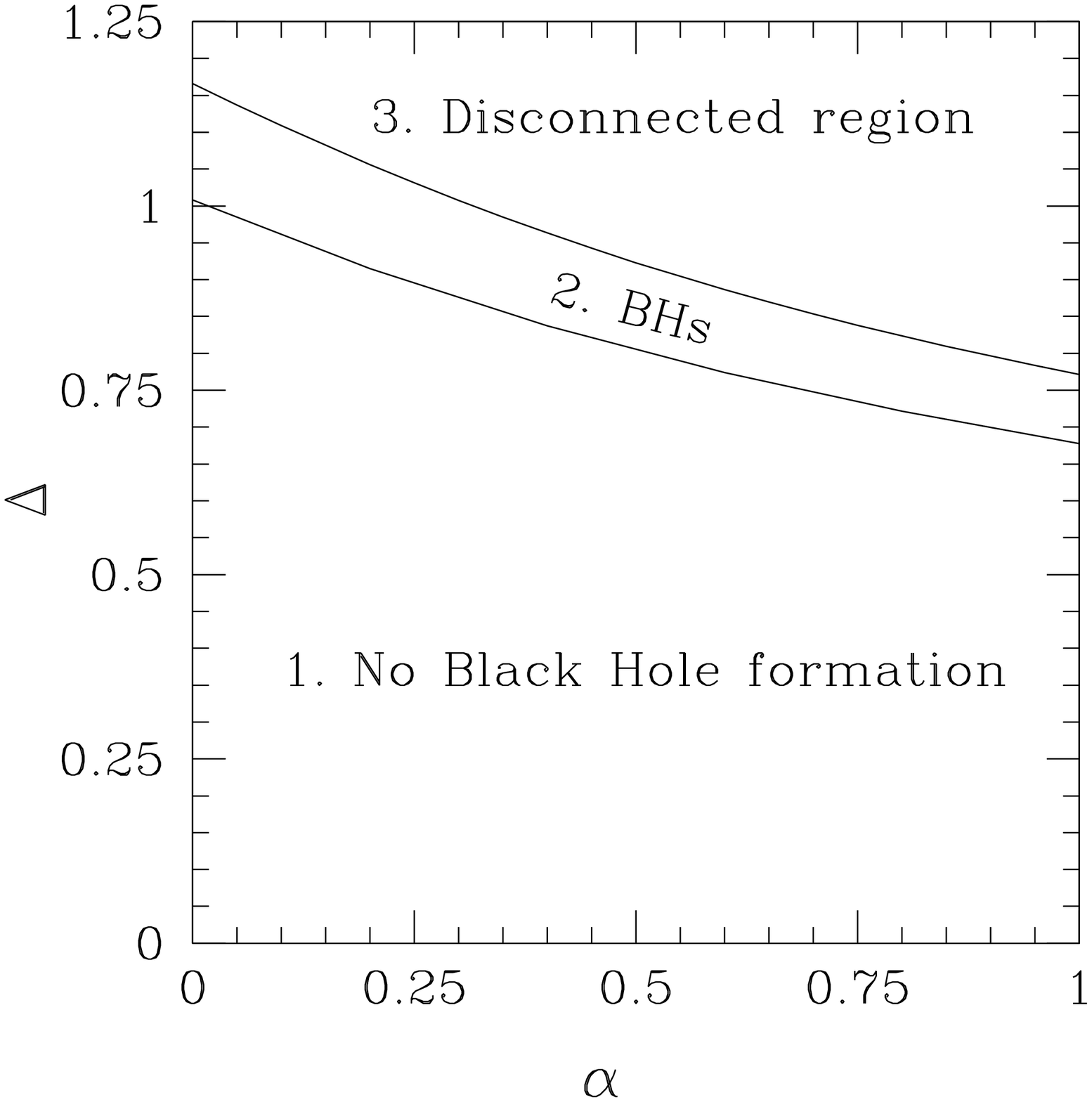,width=6.5cm}
            \psfig{file=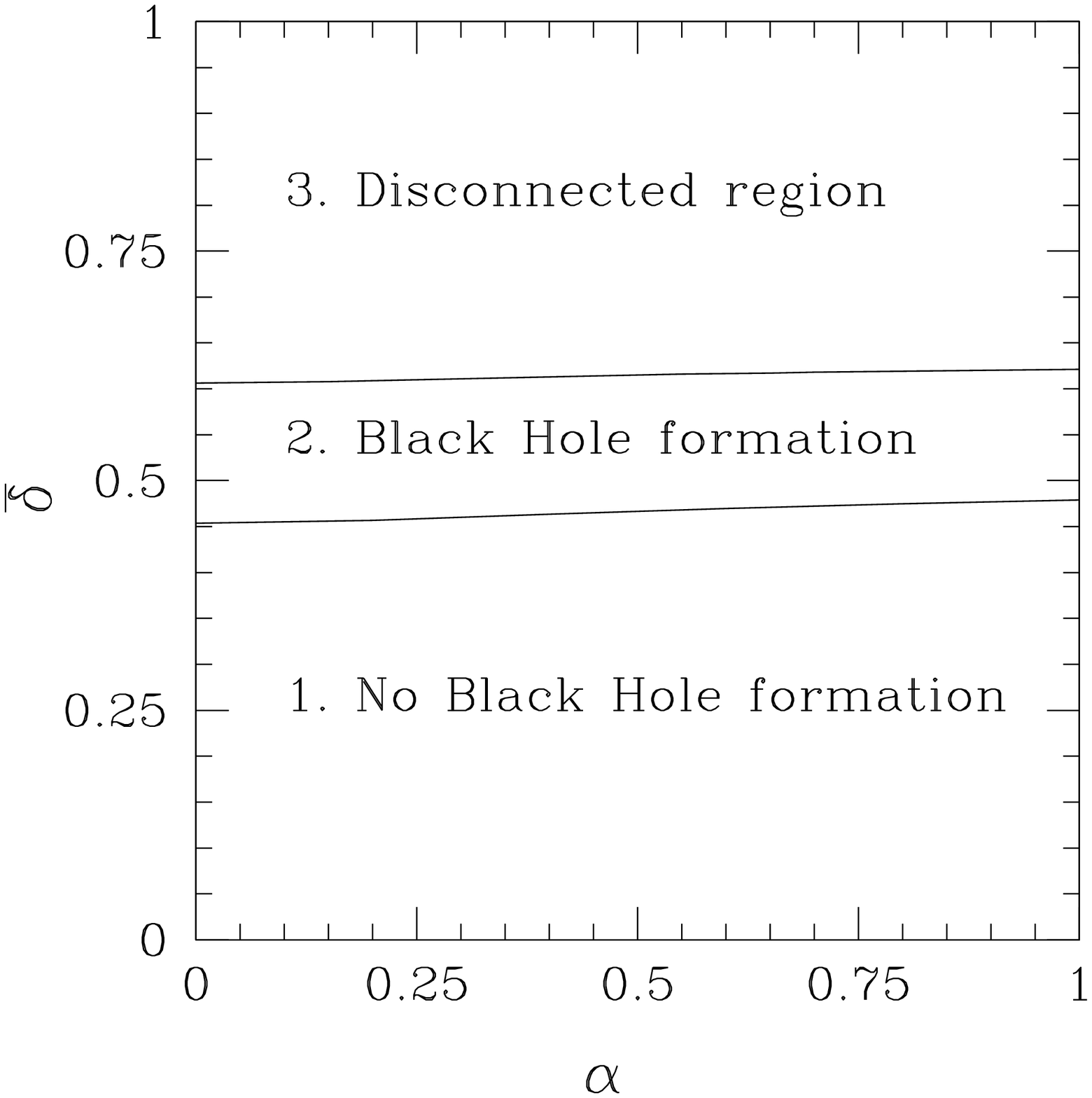,width=6.5cm}}
\caption{\label{Delta_c}\small These plots show which values of
$\alpha$, $\Delta$  and $\tilde{\delta}$ lead to black hole
formation or to an initial perturbation already
disconnected from the rest of the Universe for the case
of $K(r)$ given by expression (\ref{curv_profile}).}
\end{figure}

As a final result of these computations  we determined
the parameter ranges corresponding to black hole
formation for the curvature profiles given by
(\ref{curv_profile}) and (\ref{new_curv_param}). Figure
\ref{Delta_c} summarises the results. The left hand plot
shows the results obtained for the value of $\Delta$ and
$\alpha$ used in parametrisation (\ref{curv_profile}),
with the plane being divided into three different
regions: the first one corresponds to no black hole
formation, the second one to black hole formation and the
third one to disconnected configurations
($K(r)r^2\geq1$). It can be seen that the region of
parameters corresponding to black-hole formation is quite
narrow. 

The same results are presented in terms of
$\tilde{\delta}$ and $\alpha$ in the right hand plot of
Figure \ref{Delta_c}. This shows very weak dependence of
$\delta_c$ and $\delta_{max}$ on $\alpha$. We found that
$\tilde{\delta}_c\simeq0.45$,
$\tilde{\delta}_\mr{max}\simeq0.60$ for $\alpha=0$ and
$\tilde{\delta}_c\simeq0.47$,
$\tilde{\delta}_\mr{max}\simeq0.62$ for $\alpha=1$. 

\begin{figure}[t!]
\centerline{\psfig{file=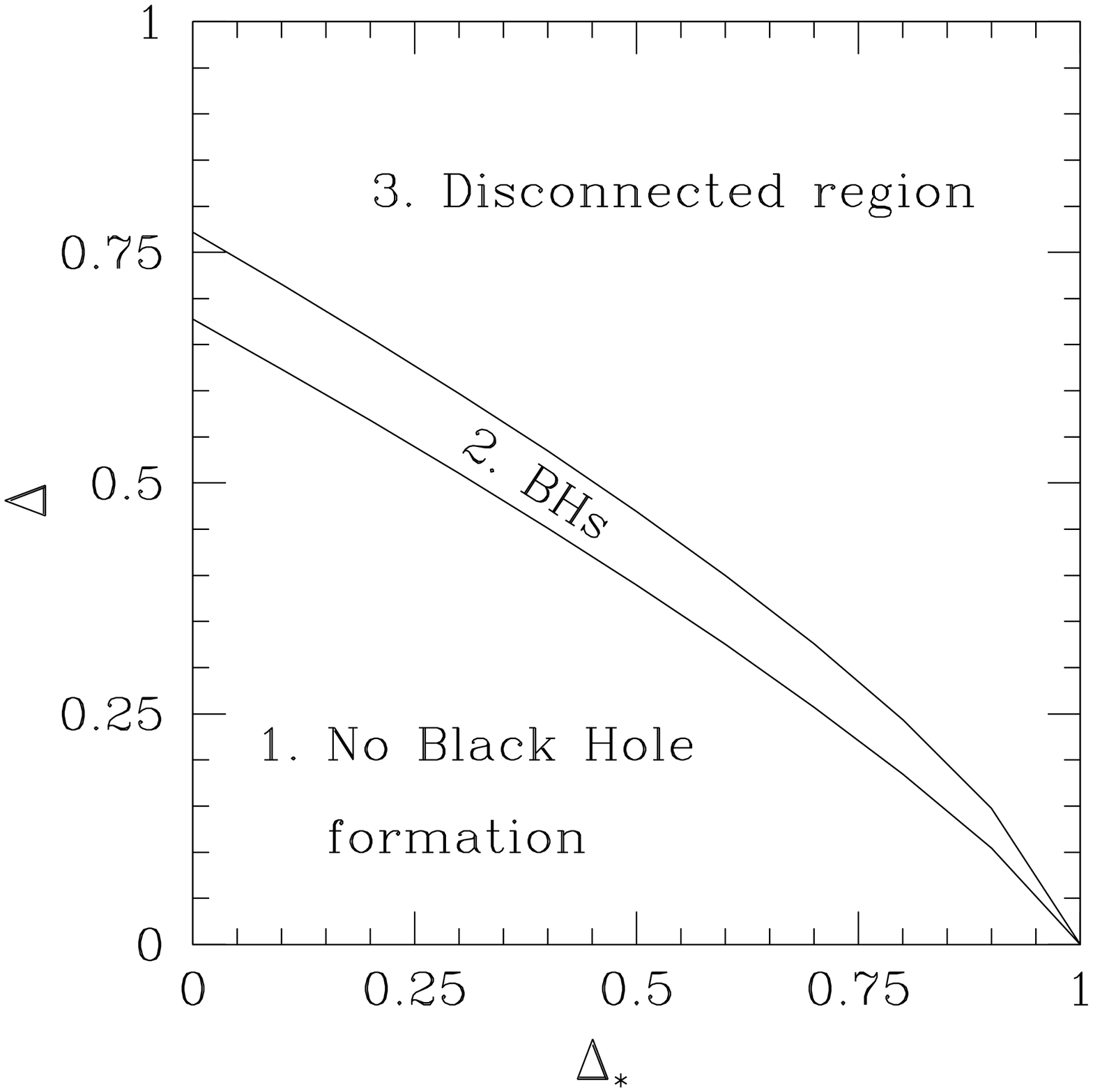,width=6.5cm}
            \psfig{file=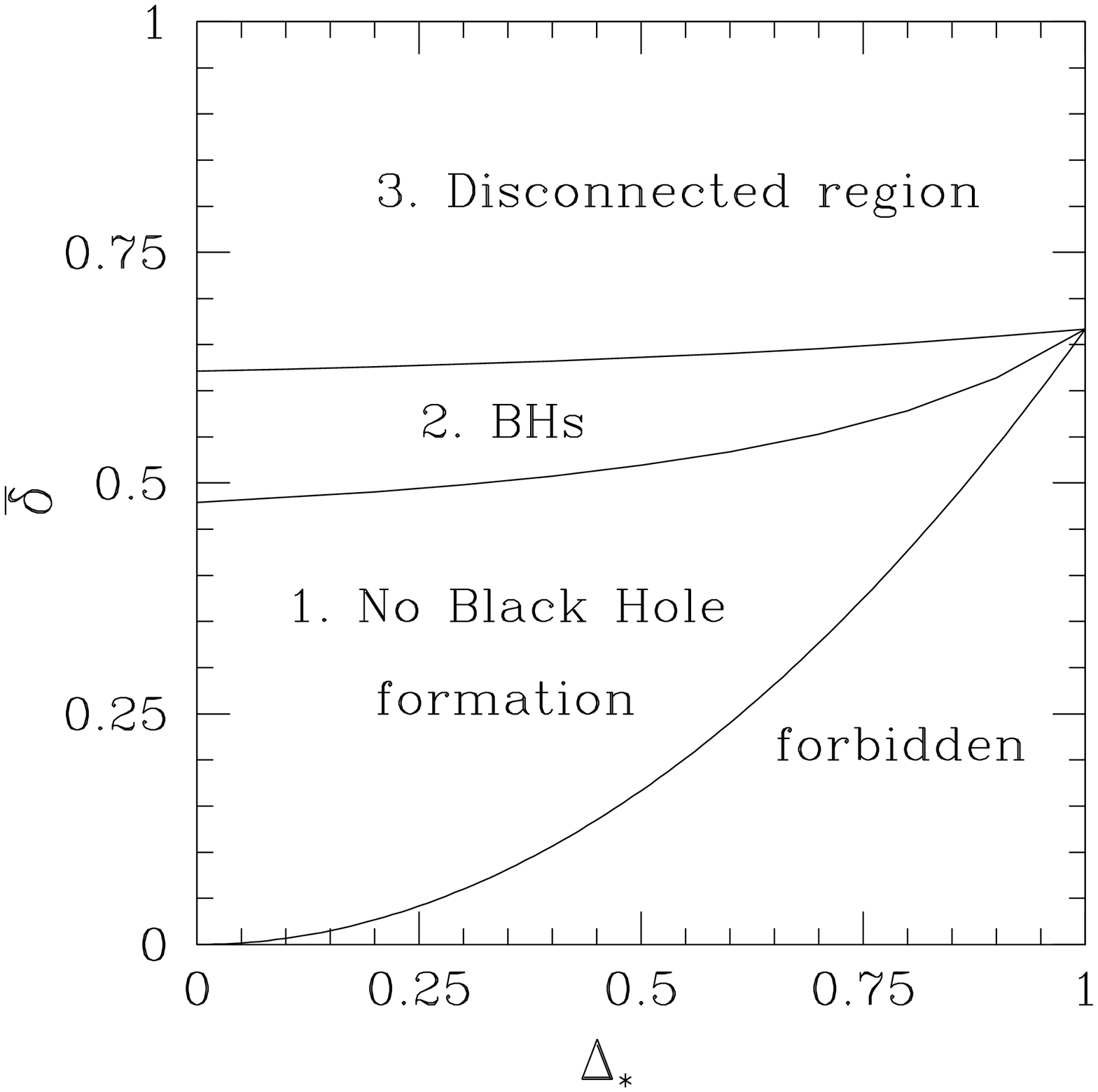,width=6.5cm}}
\caption{\label{Delta_c_new}\small These plots show the
range of parameters  $\Delta,\,\Delta_*,\,\tilde{\delta}$
giving rise to black hole formation for the case of
$K(r)$ given by (\ref{new_curv_param}).} 
\end{figure}

In Figure \ref{Delta_c_new} we present similar results
for the second curvature parametrisation given by
(\ref{new_curv_param}). The left hand plot shows the
parameter space for $\Delta$ and $\Delta_*$, and one can
see the same three regions in the left hand plot as in
Figure \ref{Delta_c}. The right hand plot shows the same
results in terms of $\Delta_*$ and $\tilde{\delta}$.

For wide range of parameters explored so far, we have not
yet seen formation of shocks as had seen in some earlier
works \cite{Nadezhin,Niemeyer1,Hawke,Musco}. Work is in
progress to investigate this further. 

\section{Conclusion}
Using the quasi homogeneous solution, we have impose
initial conditions for PBH formation by introducing a
time independent curvature profile $K(r)$, related to the
curvature perturbation $\mR$ used elsewhere in the
literature. This profile is the only source of
perturbations in the fluid quantities and those
perturbations behave as pure growing modes. We have
obtained an analytical solution of the Misner - Sharp
system of equations expressing perturbations of density
and velocity in terms of the curvature profile $K(r)$ in
the rather general case where matter in the Universe can
be treated as an arbitrary mixture of perfect fluids. 

We have performed numerical calculations for
$\gamma\,=\,1/3$ to test the self consistency of these
initial conditions for two different parameterisations of
$K(r)$. 

Specifying the initial conditions using a curvature
profile seems to be an improvement in the analysis of the
PBH formation scenario since it allows  initial
conditions for all the hydrodynamical variables relevant
for the problem to be specified in self consistent way.  

We have shown that, in agreement with \cite{Nadezhin},
the formation of PBHs requires higher amplitudes of
strong metric deviation for sharper profiles of $K(r)$.
Thus the Gaussian curvature profile, which is the least
sharp, is the most favourable for PBH formation. We found
that in this case $\delta_c\,\simeq\,0.45$ in agreement
with \cite{Musco}. The comparison of results obtained for
two different parametrisation of curvature profiles has
shown that the shape of the initial profile plays crucial
role for PBH formation. This is an important reason to
calculate the probability for having different curvature
profiles (work in progress) \cite{Hidalgo} and to link
this probability  with statistics of PBHs. Such a link
should greatly improve  the constraints on possible
cosmological models obtained from observational upper
limits on PBHs.

\vspace{0.3cm} \noindent {\bf Acknowledgements:} In the
course of this work, we have benefited from helpful
discussions with many colleagues including John Miller,
Bernard Carr, Marco Bruni, Luciano Rezzolla, Pavel
Ivanov, Andrew Liddle and Carlo Baccigalupi. 

\clearpage
\appendix
\section{Scalar curvature}

The general spherically symmetric metric in the cosmic time
coordinate is \beq ds^2 \ug -a^2dt^2 + b^2dr^2 +R^2d\Omega^2
\label{gen_metric} \eeq where the metric tensor $g_{\alpha\beta}$
is diagonal and $a$, $b$ and $R$ are, in general, functions of $r$
and $t$. The Christoffel symbols are calculated by \beq
\Gamma^\alpha_{\beta\gamma} \ug \frac{1}{2}g^{\alpha\alpha} \lp
\frac{\partial g_{\beta\alpha}}{\partial x^\gamma} +
\frac{\partial g_{\gamma\delta}}{\partial x^\beta} -
\frac{\partial g_{\beta\gamma}}{\partial x^\delta} \rp\,, \eeq
while the components of the Ricci Tensor are given by \beq
R_{\alpha\beta} \ug
\Gamma^\gamma_{\delta\gamma}\Gamma^\delta_{\alpha\beta} -
\Gamma^\gamma_{\delta\beta}\Gamma^\delta_{\alpha\gamma} +
\frac{\pt\Gamma^\gamma_{\alpha\beta}}{\pt x^\gamma} -
\frac{\pt\Gamma^\gamma_{\alpha\gamma}}{\pt x^\beta} \eeq and to
calculate the Ricci scalar we need to take into account only the
diagonal components $R_{\alpha\alpha}$.

In the background universe with the FRW metric
the metric components are therefore
\beq
a \ug 1 \quad b \ug \frac{s(t)}{\sqrt{1-Kr^2}} \quad R
\ug s(t)r\,,
\eeq
where the curvature $K$ is a constant equal to $0,\ \pm1$.
The diagonal components of the Ricci tensor are
\bea
& R_{00} \ug 3\frac{\ddot{s}}{s} \\ \nnu\\
& R_{11} \ug \frac{s\ddot{s}+2\dot{s}^2+2K}{1-Kr^2} \\ \nnu\\
& R_{22} \ug \lp s\ddot{s}+2\dot{s}+2K \rp r^2 \\ \nnu\\
& R_{33} \ug R_{22}\sin^2\theta
\eea
and these are used to calculate the scalar curvature
\beq
R \ug 6 \lp \frac{\ddot{s}}{s} + \frac{\dot{s}^2}{s^2} +
\frac{K}{s^2} \rp\,.
\label{R_flat}
\eeq
While the three curvature is obtained by removing the terms
with the time derivatives,
\beq
R^{(3)} \ug 6\frac{K}{s^2}\,.
\eeq

In the general case of the metric given by
(\ref{gen_metric}), the diagonal components of the Ricci
tensor are

\bea
& R_{00} \ug \frac{1}{a}\frac{\pt a}{\pt t} \lp
\frac{1}{b}\frac{\pt b}{\pt t} + \frac{2}{R}\frac{\pt
R}{\pt t} \rp + \frac{a}{b^2}\frac{\pt a}{\pt r} \lp
\frac{2}{R}\frac{\pt R}{\pt r} - \frac{1}{b}\frac{\pt
b}{\pt r} \rp + \nnu \\ \nnu \\
& \hspace{4.5cm} + \frac{a}{b^2}\frac{\pt^2 a}{\pt r^2} -
\frac{1}{b}\frac{\pt^2 b}{\pt t^2}
-\frac{2}{R}\frac{\pt^2 R}{\pt t^2} \\
\nnu \\
& R_{11} \ug \frac{b}{a^2}\frac{\pt b}{\pt t} \lp
\frac{2}{R}\frac{\pt R}{\pt t} - \frac{1}{a}\frac{\pt
a}{\pt t} \rp + \frac{1}{b}\frac{\pt b}{\pt r} \lp
\frac{1}{a}\frac{\pt a}{\pt r} + \frac{2}{R}\frac{\pt
R}{\pt r} \rp + \nnu \\ \nnu \\
& \hspace{4.65cm} + \frac{b}{a^2}\frac{\pt^2 b}{\pt t^2} -
\frac{1}{a}\frac{\pt^2 a}{\pt r^2}
-\frac{2}{R}\frac{\pt^2 R}{\pt r^2} \\
\nnu \\
& R_{22} \ug \frac{R}{a^2}\frac{\pt R}{\pt t} \lp
\frac{1}{b}\frac{\pt b}{\pt t} - \frac{1}{a}\frac{\pt
a}{\pt t} \rp + \frac{R}{b^2}\frac{\pt R}{\pt r} \lp
\frac{1}{b}\frac{\pt b}{\pt r} - \frac{1}{a}\frac{\pt
a}{\pt r} \rp + \nnu \\ \nnu \\
& \qquad +\lp \frac{1}{a}\frac{\pt R}{\pt t} \rp^2
- \lp \frac{1}{b}\frac{\pt R}{\pt r} \rp^2 + R
\lp \frac{1}{a^2}\frac{\pt^2 R}{\pt t^2} - \frac{1}{b^2}\frac{\pt
R}{\pt r^2} \rp + 1 \\
\nnu \\
& R_{33} \ug R_{22}\sin^2\theta
\eea
To write the scalar curvature it is
useful to define the following operators:
\beq
D_t \equiv \frac{1}{a}\frac{\partial}{\partial t}\,,
\quad
D^2_t \equiv \frac{1}{a^2}\frac{\partial^2}{\partial t^2}\,,
\quad
D_r \equiv \frac{1}{b}\frac{\partial}{\partial r}\,,
\quad
D^2_r \equiv \frac{1}{b^2}\frac{\partial^2}{\partial r^2}
\eeq
We then get
\[
\!\!\!\!\!\!\!\!\!\!\!\!\!\!\!\!\!\!\!\!
\!\!\!\!\!\!\!\!\!\!\!\!\!\!\!\!\!\!\!\! R \ug 2 \left[
\frac{D^2_t b}{b} + 2\frac{D^2_t R}{R} - \frac{D^2_r a}{a} -
2\frac{D^2_r R}{R} + \left(\frac{D_t R }{R}\right)^2 -
\left(\frac{D_r R}{R}\right)^2 + 2\frac{D_t b}{b}\frac{D_t R}{R} -
\right. \] \beq \!\!\!\!\!\!\!\!\!\!\!\!\!\!\!
\!\!\!\!\!\!\!\!\!\!\!\!\!\!\! \left. - 2\frac{D_t a}{a}\frac{D_t
R}{R} - \frac{D_t a}{a}\frac{D_t b}{b} + 2\frac{D_r b}{b}\frac{D_r
R}{R} + \frac{D_r a}{a} \frac{D_r b}{b} - 2\frac{D_r
a}{a}\frac{D_r R}{R} + \frac{1}{R^2} \right] \eeq and substituting
the expressions for the coefficients $a$, $b$ and $R$ given by
(\ref{R_def},\ref{b_def},\ref{a_def}) and expressing
$\tilde{a},\tilde{b},\tilde{R}$ in terms of $K(r)$, we get with
the use also of (\ref{dot_eps}), that the Ricci scalar is given by
\beq R \ug \lP 6\frac{\ddot{s}}{s} \lp 1 + \epsilon
\frac{3\gamma-1}{5+3\gamma}\mathcal{K} \rp + \frac{\dot{s}^2}{s^2}
\lp 1 - \epsilon \frac{2(2+3\gamma)}{5+3\gamma}\mathcal{K} \rp +
\frac{\mathcal{K}}{s^2} \rP\,, \label{general_R} \eeq where \beq
\mathcal{K} \ug K(r) + \frac{r}{3}K^\prime(r)\,. \eeq It is useful
to compare (\ref{general_R}) with (\ref{R_flat}).


\end{document}